\documentclass[superscriptaddress,twocolumn,aps,prxlife,longbibliography]{revtex4-2}

\usepackage{multirow}
\usepackage{amsmath}
\usepackage{amssymb}

\newcommand{\startsupplementnum}{%
\setcounter{table}{0}
\renewcommand{\thetable}{\thesection\arabic{table}}%
\setcounter{figure}{0}
\renewcommand{\thefigure}{\thesection\arabic{figure}}%
}

\usepackage{graphicx}
\usepackage{dcolumn}
\usepackage{bm}
\usepackage{xcolor}
\usepackage{hyperref}
\usepackage{microtype}
\usepackage{soul}

\hypersetup{
    colorlinks=true,
    linkcolor=blue!50!black,
    citecolor=blue!50!black,
    urlcolor=blue!50!black
}

\begin{document}

\title{Dynamical mean-field theory of latching Potts networks}

\author{Anindita Basu}
\email{abasu@sissa.it}
\thanks{These authors contributed equally to this work.}
\affiliation{SISSA - International School for Advanced Studies, Trieste, Italy}

\author{Carlo Orientale Caputo}
\email{corienta@sissa.it}
\thanks{These authors contributed equally to this work.}
\affiliation{SISSA - International School for Advanced Studies, Trieste, Italy}

\author{Fernando Lucas Metz}
\email{fernando.metz@ufrgs.br}
\affiliation{Physics Institute – Federal University of Rio Grande do Sul, Porto Alegre, Brazil}

\author{Alessandro Treves}
\email{ale@sissa.it}
\thanks{Corresponding author}
\affiliation{SISSA - International School for Advanced Studies, Trieste, Italy}

\begin{abstract}

Spontaneous transitions between memories, as observed in internally generated cognition such as mind-wandering and free association, require neural dynamics capable of destabilizing one attractive memory state and moving towards another. We ask what minimal representational structure is required for such transitions to emerge in random associative networks. To address this question, we study Potts associative networks with $S$ active states per unit and slow adaptive thresholds. The $S=1$ case can model a standard Hopfield-like network of binary units. We employ dynamical mean-field theory in the extensive-load regime, where the number of stored patterns scales linearly with network size and connectivity, and we focus on transitions between two condensed memories, which are singled out in the connectivity matrix. The remaining stored patterns are averaged over as quenched disorder. This construction reduces the network dynamics to a low-dimensional self-consistent description, which allows one to analyze how the overlap between two condensed memories determines adaptive transitions. In 
the finite-load
regime,  we show that transitions are governed by an effective drive term, whose sign and temporal evolution distinguish high-latching (structured), low-latching (random), and non-latching regimes. The resulting phase diagram shows that adaptation alone is not sufficient: transitions only occur within certain ranges of the overlap between the two memories. In the binary case $S=1$, in particular, no unit can be active in both memories but in different Potts states, thus strongly supressing the high-latching, structured transitions that exist for $S>1$. In the extensive-load regime, non-condensed memories generate colored fluctuations and retarded self-interactions through an effective memory kernel. We find that latching trajectories change the effective timescale of this kernel, a dynamical signature of transitions in the extensive-load regime. These results indicate that multi-state Potts-like representations are essential for structured memory-to-memory transitions, a hallmark of spontaneous thought.
\end{abstract}

\maketitle
\section{Introduction}

Associative neural network models link computation to what is available in cortical circuits, such as plain dendritic summation (unlike e.g. complex logic gates), distributed representations (unlike symbolic units) and associative plasticity. Unlike supervised learning, neuronal co-activation drives local synaptic strengthening in recurrent neural networks, which can support attractor dynamics and yield retrieval of learned patterns from partial cues. This is exemplified by Hopfield networks with two-state units \cite{Hopfield1982}, where model Hebbian learning results in symmetric couplings, defining an energy function whose minima correspond to stored memories. Extensions to graded-response units, already in \cite{Hopfield1984}, have broadened the formalism toward more realistic continuous firing rate models. The statistical-mechanics analysis of Hopfield networks with extensive loading \cite{AmitGutfreundSompolinsky1985} has characterized the retrieval, spin-glass, and paramagnetic phases, while subsequent work has shown that sparse coding can substantially increase storage capacity and reduce interference \cite{TsodyksFeigelman1988}.

The mammalian neocortex exhibits a strongly modular organization, with dense local recurrent connectivity within cortical patches and comparatively sparse long-range projections between them \cite{braitenberg2013anatomy}. This has led to consider associative network models comprised not of single neurons but of coarse-grained multi-state variables representing the collective attractors of local cortical assemblies. Potts units provide a natural implementation of this mesoscopic description: rather than representing the activity of individual neurons, each unit can be in one among $S$ local attractors, together with an inactive state corresponding to the absence of local activation. A mathematical analysis of such a multi-state associative model was first proposed by Kanter \cite{Kanter1988}, and demonstrated an almost quadratic increase in storage capacity with the number of locally available states. His treatment can be considered as an extension of that worked out for the Hopfield model which, since Kanter did not consider an inactive state, corresponds to setting $S=2$ (while e.g. the model in \cite{TsodyksFeigelman1988} corresponds to a Potts network with an inactive state and $S=1$). Subsequent work has interpreted Potts units as effective variables for cortical patches, where strong intra-module recurrence supports multiple local attractors while sparse inter-module connectivity binds them into coherent global memories \cite{Treves2005,Naim2018}.

Building on these associative memory models, it was proposed that relatively slower adaptation processes can alter network dynamics and produce, in addition to memory retrieval, spontaneous transitions between retrieved memories \cite{Treves2005}. In particular, activity-dependent fatigue within local cortical patches can progressively destabilize currently retrieved attractors, enabling transitions toward other memory states, in a regime referred to as latching. Such dynamics provide a potential mechanistic substrate for internally generated cognitive sequences, including free association and mind wandering, in which the system autonomously explores memory space without external input \cite{Treves2005,Russo2008,RussoTreves2012}.

Despite its conceptual appeal, the dynamical origin of latching remains only partially understood. In particular, it is unclear what minimal ingredients are required for adaptation to internally generate structured transitions between memories, rather than simple decay or random wandering. A key unresolved question is whether the emergence of such transitions depends fundamentally on the Potts units representing cortical patches, or they can occur also in standard models in which units represent individual neurons (which can be included in the same formalization, just setting $S=1$, see also the seminal study in \cite{horn1989neural}). In other words, can Hopfield-like networks engage in spontaneous yet structured latching (not reflecting externally imposed episodic transition as in \cite{herrmann1993neural})? 
While previous work has addressed this question numerically and through signal-to-noise analyses \cite{RussoTreves2012,Kang2017}, suggesting a negative answer, it has remained unclear whether one can draw a general and definite conclusion.

To address it, a theoretical framework is required that can separate the contributions of {\em condensed} memories (those engaging in a transition), quenched disorder, and adaptive temporal feedback in the thermodynamic limit. Dynamical mean-field theory (DMFT) provides precisely such a framework \cite{Martin1973,Coolen2001,Crisanti2018,Metz2025}. By averaging over the non-condensed patterns, DMFT reduces the high-dimensional network dynamics to an effective stochastic single-site process with self-consistently determined memory kernels and colored noise. This allows one to derive exact dynamical equations for overlaps, correlations, and response functions in the extensive-load regime \cite{Coolen2001}, going beyond finite-size simulations and heuristic arguments.

In this work, we use DMFT to study adaptive Potts associative networks with extensive loading and address two central questions. First, what minimal representational complexity is required for adaptation to generate directed memory transitions? In particular, we compare the dynamics of binary ($S=1$) and multi-state ($S>1$) networks and determine how the internal structure of the units allows for the emergence of latching dynamics. Second, what relations between memories enable transitions between them, and how are these conditions modified by the interference generated by non-condensed patterns?

We show that, under a two-condensed-memories ansatz, the dynamics reduces to only five effective classes of units whose stochastic evolution can be solved self-consistently within DMFT. Analyzing these effective dynamics reveals that non-random transitions are disabled in the binary case, $S=1$, while they occur in genuine ($S >1$) Potts networks. Our results therefore indicate that multi-state cortical representations are necessary for internally generated cognitive dynamics.

The remainder of the paper is organized as follows. In Section II, we introduce the model: an adaptive Potts associative network. We define the stored random memories, the continuous-time dynamics, and the phenomenology of high-latching and low-latching transitions. In Section III, we develop the DMFT methods, separating a finite number of condensed memories from the extensive background of non-condensed patterns, and derive the corresponding effective single-site dynamics. In Section IV, we organize results in three parts. First, we analyze the non-extensive-load limit and show how the geometry of overlap between two Potts memories determines whether high-, low-, or non-latching dynamics occurs. Second, we compare multi-state Potts networks with the binary case, showing why the transition channels required for structured latching disappear in the binary limit. Third, we study the extensive-load regime, where non-condensed memories generate colored fluctuations and retarded self-interactions, and we identify changes in the DMFT memory kernel as a dynamical signature of latching. Finally, in Sections V and VI, we discuss the implications of these results for cortical representations of semantic memories and internally generated memory sequences.

\section{Model} \label{mainmod}

We consider a network of $N$ units, where each unit models a local assembly of neurons and can be found in one of $S$ active states and one silent state denoted by $0$. The network evolves in continuous time. For each unit (indexed as $i=1,\dots,N$) and active state (indexed as $k=1,\dots,S$), we introduce a fast variable $r_i^k(t)$ representing the local field, together with a slow adaptive threshold $\theta_i^k(t)$ modeling activity-dependent fatigue of the active fraction of the local assembly, symbolically represented by a color of a Potts unit. The dynamics are given by
\begin{align}
\tau_r \dot r_i^k(t)
&= -r_i^k(t) - \theta_i^k(t) + h_i^k(t),
\label{pde_r}\\
\tau_\theta \dot\theta_i^k(t)
&= -\theta_i^k(t) + \sigma_i^k(t),
\label{koyy}\\
&\qquad i=1,\dots,N,\quad k=1,\dots,S.
\notag
\end{align}
with $\tau_r$ and $\tau_\theta$ defining the fast and slow dynamical timescales, respectively. The input field is
\begin{equation}
h_i^k(t) = \sum_{j=1}^N \sum_{l=1}^S J_{ij}^{kl} \, \sigma_j^l(t) + w\left(\sigma_i^k - \frac{1}{S} \sum_{l=1}^{S} \sigma_i^{l}\right), 
\end{equation}
where the parameter $w$ controls the local competition between active states within the same Potts unit, favoring single-state occupancy and stabilizing local categorical representations.

The instantaneous activity of each unit is obtained through a softmax nonlinearity
\begin{equation}
\sigma_i^k(t) = \frac{\exp\big(\beta r_i^k(t)\big)}
{\exp(\beta U) + \sum_{l=1}^S \exp\big(\beta r_i^l(t)\big)}, 
\label{act}
\end{equation}
where $\beta$ is analogous to an inverse temperature, and $U$ sets a global bias toward the inactive state. Larger values of $U$ increase the relative stability of the inactive state and therefore regulate the overall sparsity of the dynamics.  The dynamical equation for $\theta_i^k(t)$ can be formally integrated, 
\begin{equation}\theta_i^k(t) = \theta_i^k(0)e^{-t/\tau_\theta}
+ \frac{1}{\tau_\theta} \int_0^t dt'\, e^{-(t-t')/\tau_\theta}\, \sigma_i^k(t'),\end{equation}
showing that adaptation acts as a temporal filter over the past activity. 

The memories stored in the network are represented by $p$ random Potts patterns ${\xi_i^\mu}$ ($\mu=1,\dots,p$). Each pattern assigns to every unit $i$ either the silent state $0$ or one of the $S$ active states, so that $\xi_i^\mu \in \{ 0,1,\dots,S \}$. The sparsity of the representation, or the fraction of active units in each pattern, is controlled by the coding level $a\in(0,1)$. The pattern components are drawn with probability
\begin{equation}
\begin{aligned}
\Pr(\xi_i^\mu=0) &= 1-a,\\
\Pr(\xi_i^\mu=k) &= \frac{a}{S},
\quad k=1,\dots,S.
\end{aligned}
\end{equation}

Thus, each stored pattern contains on average $Na$ active units, whose active states are uniformly distributed among the $S$ Potts colors.
We assume that at any given time the network has macroscopic
overlap with a finite number $B$ of condensed patterns ($\mu=1,\dots,B$). For a specific realization of stored patterns, we explicitly parametrize the correlations among the $B$ patterns for a given unit $i$, whereas we treat the remaining $p-B$ non-condensed patterns as statistically independent. To investigate how the similarity between the stored memories affects latching, we will focus on the case $B=2$ and use as control parameters the number $C_1$ of units that are active in the same Potts state in both patterns, and the number $C_2$ of units that are active in different Potts states. 

The total number of patterns scales as $p=\alpha N$, where $\alpha$ is the loading parameter (the natural scaling would be with the number of connections per unit, which itself may or may not be considered to scale with $N$ \cite{mari1998modeling}, but we leave the case of a non-fully connected network to future studies). 
The synaptic tensor $J_{ij}^{kl}$ is defined by the Potts Hebbian learning rule
\begin{equation}
J_{ij}^{kl} \;=\; \frac{1-\delta_{ij}}{N\,\tilde{a}}\sum_{\mu=1}^p 
\eta_{ik}^{\mu} \eta_{jl}^{\mu} ,
\label{eq:J}
\end{equation}
where
\begin{equation}
\eta_{ik}^{\mu} = \delta_{\xi_i^\mu,k}-\frac{a}{S}, 
\quad
\tilde{a}=a\left(1-\frac{a}{S}\right)
\end{equation}
with $i,j=1,\dots,N$ and $k,l=1,\dots,S$.
The factor $1-\delta_{ij}$ removes self-couplings, while the centering terms $-a/S$ subtract the mean activity of each active state. 

\subsection{Phenomenology of latching}

Under suitable conditions, the network exhibits latching dynamics, in which activity transiently remains near a retrieved memory before transitioning to another attractor. These transitions are driven by the slow adaptive field introduced above. During the retrieval of a given memory, the adaptive thresholds associated with its active local states gradually increase over the timescale $\tau_\theta$, reducing the stability of the corresponding attractor. Once this attractor becomes sufficiently destabilized, the network relaxes toward a different memory configuration, where the same mechanism repeats. The resulting dynamics consists of a sequence of metastable memory states whose dwell times are controlled by the interplay between adaptation, recurrent reinforcement, and noise. As observed in previous work~\cite{Russo2008}, depending on the overlap between stored memories, the network displays qualitatively different classes of transitions. These preliminary observations motivate the dynamical mean-field analysis developed in the following sections.

\begin{figure}[t]
\centering
\includegraphics[width=\linewidth]{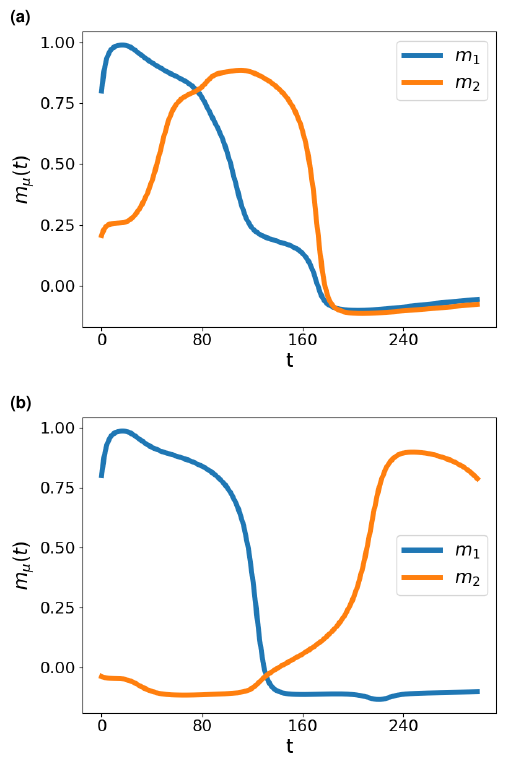}

\caption{Numerical simulation results for the overlap dynamics during high- and low-latching transitions in a network with $N=1000$ units, $S=3$ Potts states, and $p=2$ stored patterns. The network is initialized close to the first pattern. The other model parameters are (see section \ref{mainmod}): \(a=0.25\), \(U=0.3\), \(\beta=5\), \(\tau_r=10\), \(\tau_\theta=100\) and \(w=0.8\). (a) High-overlap latching: \(X=0.8 \) and \(Y=0.4\). (b) Low-overlap latching: \(X=0.05\) and \(Y=0.8\).}
\label{fig: high_low}
\end{figure}

When only two patterns are stored in the connectivity matrix of a large network, numerical simulations reveal qualitatively distinct transition regimes:

\begin{itemize}

\item \textbf{High-overlap latching:} transitions in which the overlap with the second memory increases while the first memory is still substantially active (see Fig. \ref{fig: high_low}(a)). These events typically occur between partially overlapping memories.

\item \textbf{Low-overlap latching:} transitions in which the first memory is inactivated before the second memory becomes active (see Fig. \ref{fig: high_low}(b)). These events are more common for weakly correlated or uncorrelated memories and proceed through an intermediate low-activity network state.

\item \textbf{Pathological latching}: transitions in which the network alternates repeatedly
between two highly correlated memories. This regime is termed pathological latching in \cite{Russo2008}.%

\end{itemize}
Examples of high- and low-overlap latching are shown in Fig. \ref{fig: high_low}, which illustrates the dynamics of the network overlaps with the first and second memories for different values of $C_1$ (units active in the same Potts state in the two memories) and $C_2$ (units active in different Potts states). 
To investigate how the overlap between the patterns influences latching dynamics, it is convenient to introduce \cite{Russo2008}
\begin{equation}
X = \frac{C_1}{C_1+C_2} \quad \text{and} \quad Y = \frac{C_1+C_2}{Na},
\end{equation}
where $X$ is the fraction of units active in the same Potts state, while $Y$ is 
the total fraction of units active in both memories. High-overlap latching transitions occur for larger values of $X$, when the two patterns share many units active in the same Potts state. 

This phenomenology remains essentially the same when introducing models of slow or fast inhibition \cite{Russo2008,RussoTreves2012} or a combination of the two, as shown by extensive simulations (Kwang Il Ryom, unpublished studies and \cite{ryom2023statistical}): even though the type of {\em phase} transitions between the different regimes may vary, the types of {\em latching} transitions are those described above,

Latching from one memory pattern to the next relies on the ability of each unit to represent multiple active states. This multi-state structure appears to be essential: previous numerical studies have shown that networks with binary units do not sustain robust memory transitions \cite{Treves2005,Kang2017}.
Although adaptive thresholds destabilize the retrieved memory, they do not efficiently drive transitions toward alternative attractors in binary-state networks. These observations motivate the development of a theoretical framework capable of identifying the dynamical mechanisms underlying memory transitions.

\section{Dynamical mean-field theory}

To study the dynamics in the extensive-load regime, we employ dynamical mean-field theory (DMFT), which determines the generating functional of dynamical path-probabilities after averaging over the disorder associated with the non-condensed patterns. In the limit $N \rightarrow \infty$, this procedure reduces the original dynamics of $N$ coupled units to a set of effective single-site stochastic equations that depends self-consistently on the overlaps with the condensed patterns, as well as on correlations and response functions. The resulting equation also depends explicitly on the specific realization of the finite set of condensed patterns, which allows the dynamical equations to be further reduced to a finite number of effective site classes. In appendix \ref{app: dis_average}, we develop the DMFT for the present model and derive the effective stochastic equations.

The generating functional of the dynamics is defined as
\begin{equation}
\mathcal{Z}[\boldsymbol{\psi}]
=
\Bigg\langle
\exp{\left(
\int_{0}^{T} dt \sum\limits_{i = 1}^N \sum\limits_{k=1}^S
\psi_i^k(t) \sigma_i^k(t) \right)  } 
\Bigg\rangle,
\end{equation}
where $\langle \dots \rangle$ denotes the average with respect  
to the path-probability $\mathcal{P}\left[\boldsymbol{r}\right]$ 
of the microscopic fields $\{ r_{i}^{k}(t) \}$ over the time interval $t\in[0,T]$. The 
generating functional encodes the statistics of the time-dependent network 
activity, since  derivatives of $\mathcal{Z}[\boldsymbol{\psi}]$ with 
respect to the source fields $\psi_{i}^{k}(t)$ generate the correlation functions of $\sigma_i^k(t)$. In 
particular, the average single-site activity and the two-time activity correlation 
are given by
\begin{equation}
\langle \sigma_i^k(t) \rangle = \frac{\delta \mathcal{Z} }{\delta \psi_{i}^{k}(t) }  \Big{|}_{\psi=0} 
\end{equation}
and
\begin{equation}
\langle \sigma_i^k(t) \sigma_i^l(t') \rangle = \frac{\delta^{2} \mathcal{Z} }
{\delta \psi_{i}^{k}(t) \delta \psi_{i}^{l}(t') }\Big{|}_{\psi=0} .   
\end{equation}

In appendices \ref{app: dis_average} and \ref{app: cavity_method}, we show that, in 
the limit $N \rightarrow \infty$, the original network
dynamics factorizes into $M(B,S)$ effective stochastic single-site equations
\begin{equation}
\label{eq: single site eq}
\begin{aligned}
\tau_r \dot r_g^k(t)
&= -r_g^k(t) - \Omega_g^k(t)
 + \sum_{\mu=1}^{B}\eta_{gk}^\mu m^\mu(t)\\
&\quad + \frac{1}{\tilde a}
 \int_0^t dt'\sum_{l=1}^{S}
 C_{kl}K(t,t')\sigma_g^l(t')
 + \phi_g^k(t).
\end{aligned}
\end{equation}
where
\begin{equation}
\Omega_{g}^{k}(t) = \theta_{g}^{k}(t) - 
w \Big( \sigma_{g}^k(t) - \frac{1}{S} \sum_{l=1}^{S} \sigma_{g}^{l}(t) \Big).
\label{rewd}
\end{equation}
The index \(g=1,\ldots,M(B,S)\) labels distinct single-site pattern configurations 
(see appendix \ref{app: 2cond_K} for an explicit expression for $M(B,S)$).
The effective stochastic dynamics depends on the average overlap
with the condensed patterns
\begin{equation}
\begin{aligned}
m^\mu(t)
&= \lim_{N\to\infty}
\frac{1}{N\tilde a}
\sum_{i=1}^N \sum_{k=1}^S
\eta_{ik}^\mu \langle \sigma_i^k(t) \rangle,\\
&\qquad\qquad\qquad \mu=1,\dots,B.
\end{aligned}
\end{equation}

the two-time correlation of activities
\begin{equation}
q_{kl}(t,t')
= \lim\limits_{N \rightarrow \infty}
\frac{1}{N}
\sum\limits_{i=1}^N
\langle \sigma_i^k(t)\sigma_i^l(t') \rangle,
\end{equation}
and the response of the mean activity 
\begin{equation}
G_{kl}(t,t')
= \lim\limits_{N \rightarrow \infty}
\frac{1}{N}
\sum\limits_{i=1}^N
\frac{\delta \langle \sigma_i^k(t) \rangle}
{\delta \tilde\theta_i^l(t')}\Big{|}_{\tilde\theta=0}
\end{equation}
at time $t$ due to an external perturbation $\{\tilde\theta_i^k(t')\}$ at time $t' < t$.
Equation (\ref{eq: single site eq}) contains three distinct contributions beyond the local relaxation and adaptation terms. The condensed memories generate a deterministic driving field through the overlaps $m^{1}(t), \dots,m^{B}(t)$, while the non-condensed disorder induces both a retarded self-interaction kernel and a temporally correlated Gaussian noise. 
The memory kernel is given by
\begin{equation}
\label{eq: memory kernel}
K(t,t')
=
\alpha
\left[
\tilde G
(\mathbb{I}-\tilde G)^{-1}
\right](t,t'),
\end{equation}
while $\phi_g^k(t)$ is a Gaussian noise with zero mean and covariance
\begin{equation}
\label{eq: Gaussian noise}
\begin{aligned}
\langle \phi_g^k(t)\phi_{g'}^l(t') \rangle
&= \delta_{gg'}D_{kl}(t,t')\\
&= \frac{\alpha\,\delta_{gg'}C_{kl}}{\tilde a}
\Bigl[
(\mathbb I-\tilde G)^{-1}
\tilde q\\
&\qquad {}\times
(\mathbb I-\tilde G^\top)^{-1}
\Bigr](t,t').
\end{aligned}
\end{equation}
with $\mathbb{I}(t,t')=\delta(t,t')$. The operators $\tilde G(t,t')$ and $\tilde q(t,t')$ are obtained from the correlation $q_{kl}(t,t')$ and response $G_{kl}(t,t')$ as follows
\begin{equation}
\tilde q(t,t')
=
\frac{1}{\tilde{a}}
\sum_{kl}
C_{kl}\,
q_{kl}(t,t'), 
\label{pq1}
\end{equation}
\begin{equation}
\tilde G(t,t')
=
\frac{1}{
\tilde{a}}
\sum_{kl}
C_{kl}\,
G_{kl}(t,t'),
\label{pq2}
\end{equation}
where 
\begin{equation}
C_{kl}
=
-\frac{a^2}{S^2}
+
\frac{a}{S}\delta_{kl}.
\end{equation}
Thus, the effect of quenched disorder is to generate both colored noise 
and temporal feedback, whose structure is entirely determined by the macroscopic observables.

\subsection{Classification of single-site dynamics with two condensed patterns}
\label{uiui}

For \(B=2\) and \(S\ge 2\), the two condensed patterns generate only five different
configurations of the drift term in Eq. (\ref{eq: single site eq}), up to a 
relabeling of the active Potts colors.
Let $x,y \in \{ 1,\dots,S \}$, with $x \neq y$, denote two distinct active 
states of a condensed pattern component, and let $s$ denotes the inactive 
state. If both pattern
components $(\xi_{i}^{1},\xi_{i}^{2})$ are active at a site, they can either 
have the same color or different colors, giving the 
configurations \(xx\) and \(xy\), respectively. If only one component is 
active, it can either be the first or the second pattern, we call these configurations \(xs\) and \( sy\). Finally, if both components
are inactive, the configuration is \(ss\). Thus, the relevant configurations of the 
condensed patterns in Eq. (\ref{eq: single site eq}), for $B=2$, are represented by $g \in \{    
xx,\; xy,\; sy,\; xs,\; ss \}$.
The normalized weights of each configuration $g$ are defined as
\begin{align}
c_{xx}
&=
\lim_{N\to\infty}
\frac{1}{N}
\sum_{i=1}^N
(1-\delta_{\xi_i^{1},0})
\delta_{\xi_i^1,\xi_i^{2}},
\\[2mm]
c_{xy}
&=
\lim_{N\to\infty}
\frac{1}{N}
\sum_{i=1}^N
(1-\delta_{\xi_i^1,0})
(1-\delta_{\xi_i^2,0})
(1-\delta_{\xi_i^1,\xi_i^2}),
\\[2mm]
c_{xs}
&=
\lim_{N\to\infty}
\frac{1}{N}
\sum_{i=1}^N
\delta_{\xi_i^{2},0}
(1-\delta_{\xi_i^1,0}),
\\[2mm]
c_{sy}
&=
\lim_{N\to\infty}
\frac{1}{N}
\sum_{i=1}^N
\delta_{\xi_i^{1},0}
(1-\delta_{\xi_i^2,0}),
\\[2mm]
c_{ss}
&=
\lim_{N\to\infty}
\frac{1}{N}
\sum_{i=1}^N
\delta_{\xi_i^{1},0}
\delta_{\xi_i^{2},0}.
\end{align}
The macroscopic observables are obtained by averaging the dynamical 
variables over the effective stochastic dynamics associated with each
configuration $g$, weighted by the corresponding $c_g$.
In appendix \ref{app: 2cond_K}, we derive a general expression
for the total number of configurations $M(B,S)$ for arbitrary $B$ and $S$.

The dynamical variables $(r_{g}^{k}(t),\sigma_{g}^{k}(t),\theta_{g}^{k}(t))$ are
conditioned to the pattern configurations $g = (\xi_{i}^{1},\xi_{i}^{2})$. For instance, $r_{g}^{k}$ denotes the field at color $k$ for a fixed configuration $g$. We also 
note that one must 
distinguish between the dynamical variables $r_{g}^k$ associated with different colors.
The color $k$ can either coincide with one of the active colors present in the configuration $g$, or correspond to an active color different from those appearing in $g$. In appendix \ref{app: 2cond_K}, we explain how to enumerate all distinct dynamical 
variables. For $B=2$, this procedure
yields ten distinct equations for the dynamical fields.

For $B=2$, the dynamics depends on the overlap structure between the two condensed patterns. 
This overlap geometry is fully specified by $\frac{C_1}{N}\simeq c_{xx}=aXY$ 
and $\frac{C_2}{N}\simeq c_{xy}=a(1-X)Y$, corresponding, respectively, to the fraction of units active in the same Potts state in both memories and the fraction of units active in different Potts states. 
The weights of the remaining configurations can be obtained similarly,
\begin{align}
c_{xs}&=c_{sy} = a(1-Y),
\\
c_{ss} &= 1-2a+aY.
\end{align}

\subsection{Comparing DMFT and finite-size simulations} 

\begin{figure}[t]
\centering
\includegraphics[width=\linewidth]{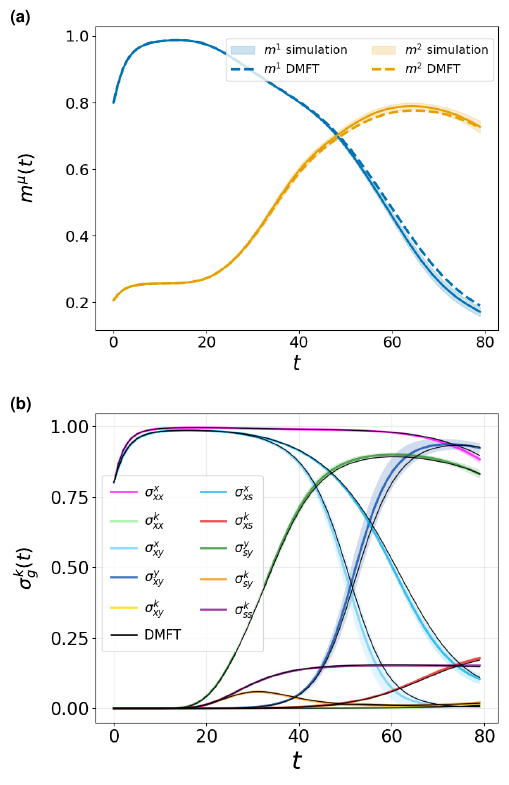}

\caption{
Finite-size simulations vs.\ DMFT at \(N=1000\), \(S=3\),
\(p=100\) (\(\alpha=0.1\)), \(a=0.25\), \(U=0.1\),
\(\beta=5\), \(\Delta t=1\), \(\tau_r=10\),
\(\tau_\theta=25\), \(w=0.8\), \(X=0.8\), \(Y=0.4\).
Top: overlaps \(m^1(t)\) and \(m^2(t)\).
Bottom: activations \(\sigma_g^k\) for each of the five groups
\((xx,xy,xs,sy,ss)\) and each of the possible colours within the groups.
Shaded bands show \(\pm1\) standard deviation over \(15\) disorder
realisations; DMFT predictions are shown as dashed/black curves.
}
\label{fig: overlap alpha=0.1}
\end{figure}

\begin{figure}[t]
\centering
\includegraphics[width=\linewidth]{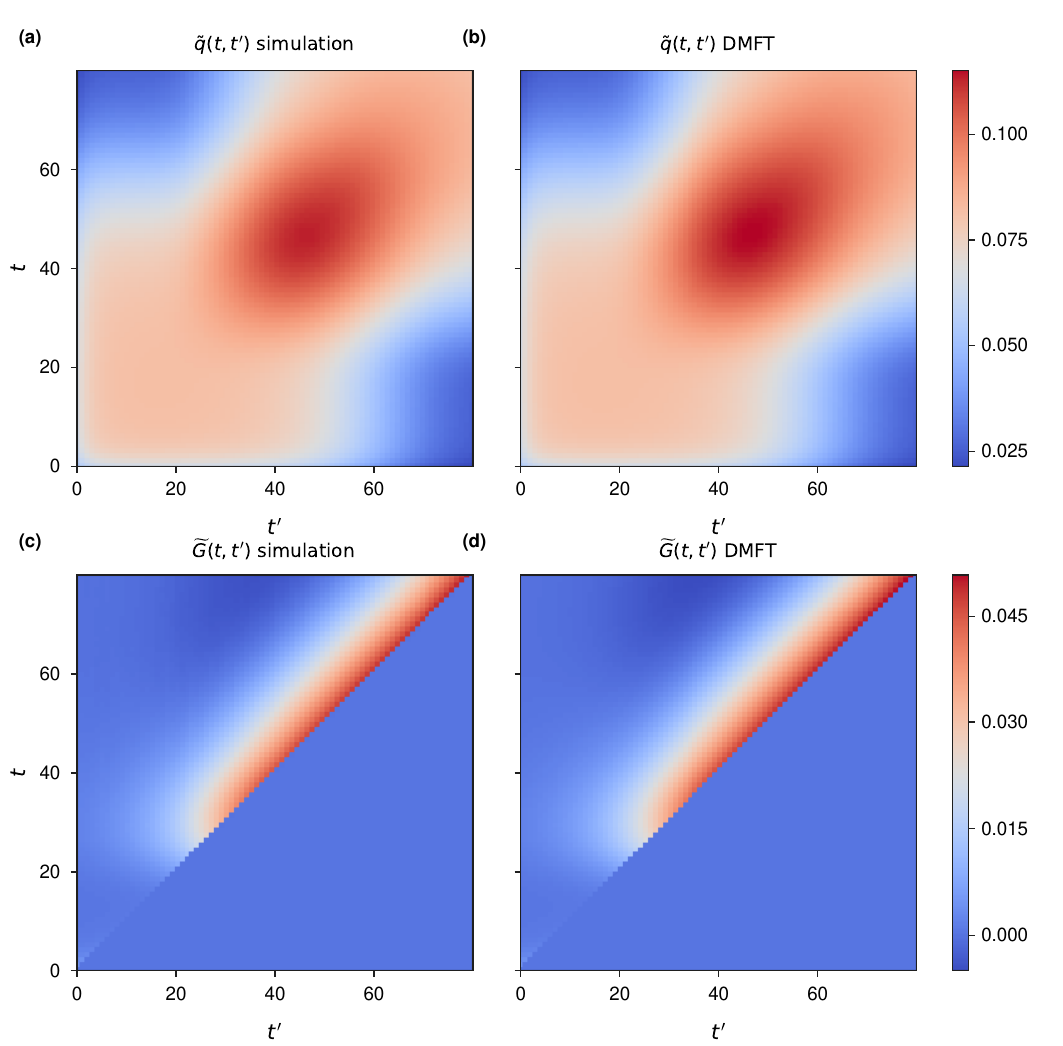}

\caption{
Comparison between DMFT and finite-size simulations for the two-time
correlation $\tilde q(t,t')$, Eq.~(\ref{pq1}) (top row),
and the response function $\tilde G(t,t')$, Eq.~(\ref{pq2})
(bottom row). The left column shows finite-size simulations,
while the right column shows the corresponding DMFT predictions.
Parameters: same as in Fig.~\ref{fig: overlap alpha=0.1}.
}
\label{fig:C_R_sim_vs_dmft}
\end{figure}

Here, we compare the predictions of DMFT with 
finite-size simulations for $B=2$ and $\alpha > 0$. The simulation results
are  obtained by directly integrating Eqs. (\ref{pde_r}) and (\ref{koyy}) for large values of
$N$.
Figure \ref{fig: overlap alpha=0.1} shows the dynamics of the 
macroscopic overlaps, $m^{1}(t)$ and $m^{2}(t)$, together with the average 
microscopic activity \(\sigma_g^k\) associated with each configuration $g$. 
Figure \ref{fig:C_R_sim_vs_dmft} displays the projected correlation
\(\tilde q(t,t')\) and response \(\tilde G(t,t')\) functions in the two-time plane $(t',t)$.
In both figures, DMFT and simulations show excellent agreement, confirming
the exactness of DMFT in the limit $N \rightarrow \infty$.

\section{Results}

\subsection{Latching transitions for $\alpha=0$}

We first analyze the finite-load regime ($\alpha=0$), in which the disorder induced by 
the non-condensed patterns is absent. In this regime, and for a fixed realization
of the condensed patterns, the DMFT equations
become  deterministic: the field $r_{g}^{k}(t)$ depends only on quantities evaluated
at the previous time, so the dynamical equations can  be integrated 
numerically without using the more elaborate algorithm discussed in appendix \ref{Numerics}.

\subsubsection{Transition channels and effective drive}

The transition from the first memory toward the second is controlled primarily by the field $r_{g}^{y}(t)$ at color $y$ of the second condensed pattern of the $xy$ and $sy$ units.
These are precisely the units that must change activity during a latching event. When the network is initially aligned with the first memory, the $sy$ units are silent, while the $xy$ units are active in the color associated with the first memory. During a latching transition, the $y$ component of these units must become active.
Hence, from Eq. (\ref{eq: single site eq}), we can derive the dynamical equations for the
relevant fields (assuming $\tau_r=1)$
\begin{align}
\dot r_{xy}^{\,y}
&= M_{12}(\{\sigma_g\})
   -\theta_{xy}^{\,y}-r_{xy}^{\,y}
\notag\\
&\quad + k_{oy}^{\,y}
\Bigl[(1-X)Y\sigma_{xy}^{\,y}
      +(1-Y)\sigma_{sy}^{\,y}\Bigr]
\notag\\
&\quad + \frac{w}{S}
\Bigl[(S-1)\sigma_{xy}^{\,y}
      -\sigma_{xy}^{\,x}
      -(S-2)\sigma_{xy}^{\,k}\Bigr],
\\[1ex]
\dot r_{sy}^{\,y}
&= M_{12}(\{\sigma_g\})
   -\theta_{sy}^{\,y}-r_{sy}^{\,y}
\notag\\
&\quad + k_{oy}^{\,y}
\Bigl[(1-X)Y\sigma_{xy}^{\,y}
      +(1-Y)\sigma_{sy}^{\,y}\Bigr]
\notag\\
&\quad + \frac{w}{S}
\Bigl[(S-1)\sigma_{sy}^{\,y}
      -(S-1)\sigma_{sy}^{\,k}\Bigr],
\label{twokeyeqs}
\end{align}
with
\begin{equation}
k_{oy}^{\,y}
= \frac{a^2/S^2+(1-a/S)^2}{1-a/S}.
\end{equation}
The term $M_{12}$ contains the contribution from all remaining pattern configurations and
acts as an effective transition drive between the two memories. Its full expression
is given in appendix \ref{app: xyy syy equations}.

Numerical simulations show that, during latching events, the activities associated with colors unrelated to the condensed memories remain strongly suppressed (see Fig. \ref{fig: overlap alpha=0.1}). Neglecting these activities in the full expression for $M_{12}$
yields the approximation
\begin{equation}
\begin{aligned}
M_{12}(\{\sigma_g\})
&\simeq (1-2a/S)XY\sigma_{xx}^x\\
&\quad - \frac{2a}{S}
\Bigl[
Y(1-X)\sigma_{xy}^x
+(1-Y)\sigma_{xs}^x
\Bigr].
\end{aligned}
\label{M12}
\end{equation}
This expression immediately reveals the geometric origin of the transition mechanism. The positive contribution to the transition drive is proportional to $XY\sigma_{xx}^{x}$, showing that adaptive transitions are promoted both by the fraction $X$ of units active in the same color in the two memories and by the fraction $Y$ of units jointly active in both memories.

In the same vein, we can derive the expression for the field that pushes the network towards the first memory: $M_{21}$. This is the common field of the units $xs$ and $xy$ at color $x$ of the first pattern:
\begin{equation}
\begin{aligned}
M_{21}(\{\sigma_g\})
&\simeq (1-2a/S)XY\sigma_{xx}^x\\
&\quad - \frac{2a}{S}
\Bigl[
Y(1-X)\sigma_{xy}^y
+(1-Y)\sigma_{sy}^y
\Bigr].
\end{aligned}
\label{M21}
\end{equation}

\subsubsection{High-latching and low-latching regimes}

To characterize the high- and low-latching regimes (see Fig.\ref{fig: high_low}) quantitatively, we introduce observables associated with the effective transition drive $M_{12}$.

The onset of high-latching is monitored through the time-averaged effective drive
\begin{align}
\overline{M}_{\rm HIGH}
=
\frac{1}{\Delta t}
\int_0^{\Delta t}
dt \;
M_{12}(\{\sigma_g  \}),
\end{align}
evaluated over a transient time window $\Delta t = \tau_\theta$. Positive values of $\overline{M}_{\rm HIGH}$ indicate that the effective transition drive remains sufficiently strong during the destabilization of the first memory.

Low-latching events exhibit a more subtle dynamics. Simulations show that they are typically associated with two successive increases in the effective drive, with the second increase occurring within a time window approximately between $2\tau_\theta$ and $3\tau_\theta$. In particular, when this second increase is absent, LOW-latching transitions are no longer observed.
Therefore, we define the observable
\begin{equation}
\overline{M}_{\rm LOW}
=
\frac{1}{\tau_{\theta}}
\int_{2\tau_{\theta}}^{3\tau_{\theta}}
dt\;
\frac{dM_{12} (\{  \sigma_g \} )}{dt} ,
\end{equation}

which measures the average growth of the transition drive during the time interval where the 
second increase emerges.

\subsubsection{Pathological latching regimes}
When the overlap between the two memories is large enough, the network can
alternate between them either a finite or an infinite number of times; we call
these the \emph{transient} and the \emph{sustained} pathological latching
regimes, respectively. Both are made possible by the presence of a limit-cycle
attractor. Depending on the overlap between the two memories, three scenarios
can occur:
\begin{itemize}
    \item an unstable limit cycle coexists with a stable fixed point
          corresponding to the silent state;
    \item the network is bistable: the limit cycle and the silent fixed point
          are both stable;
    \item the silent fixed point loses its stability, and the limit cycle is
          the only remaining attractor.
\end{itemize}
In the first scenario the network can display transient pathological latching:
it alternates between the two memories a finite number of times and eventually
decays to the silent fixed point. In the second scenario the outcome depends on
the initial condition: the network can either settle on the limit cycle,
alternating indefinitely between the two memories, or decay to the silent
state. In the third scenario the network can only end up in the sustained
pathological latching regime, since the limit cycle is the only stable
attractor of the system.
 
We find that the boundary separating the high-latching regime from the
pathological ones is captured by the maximum, over the time interval
$[\tau_{\theta},2\tau_{\theta}]$, of the effective transition drive
$\hat{M}_{21}=M_{21}-\theta_{xs}^{x}$, where $M_{21}$ is given by
Eq.~\eqref{M21}. As soon as the drive $M_{21}$ that pushes the network back towards the
first memory exceeds the adaptation of the units that code for it, $\theta_{xs}^{x}$, the network can return to the first memory. So the boundary between high latching region and pathological latching region is determined by the change in sign (from negative to positive) of $\hat{M_{21}}$.

The two boundaries separating the three scenarios described above can also be
located numerically: the first is the locus of points at which the limit cycle
becomes stable, the second the locus at which the silent state loses its
stability. The numerical details are given in Appendix \ref{app: limit cycle}.

\begin{figure*}[!tp]
\centering
\includegraphics[
  width=\textwidth,
  height=0.75\textheight,
  keepaspectratio
]{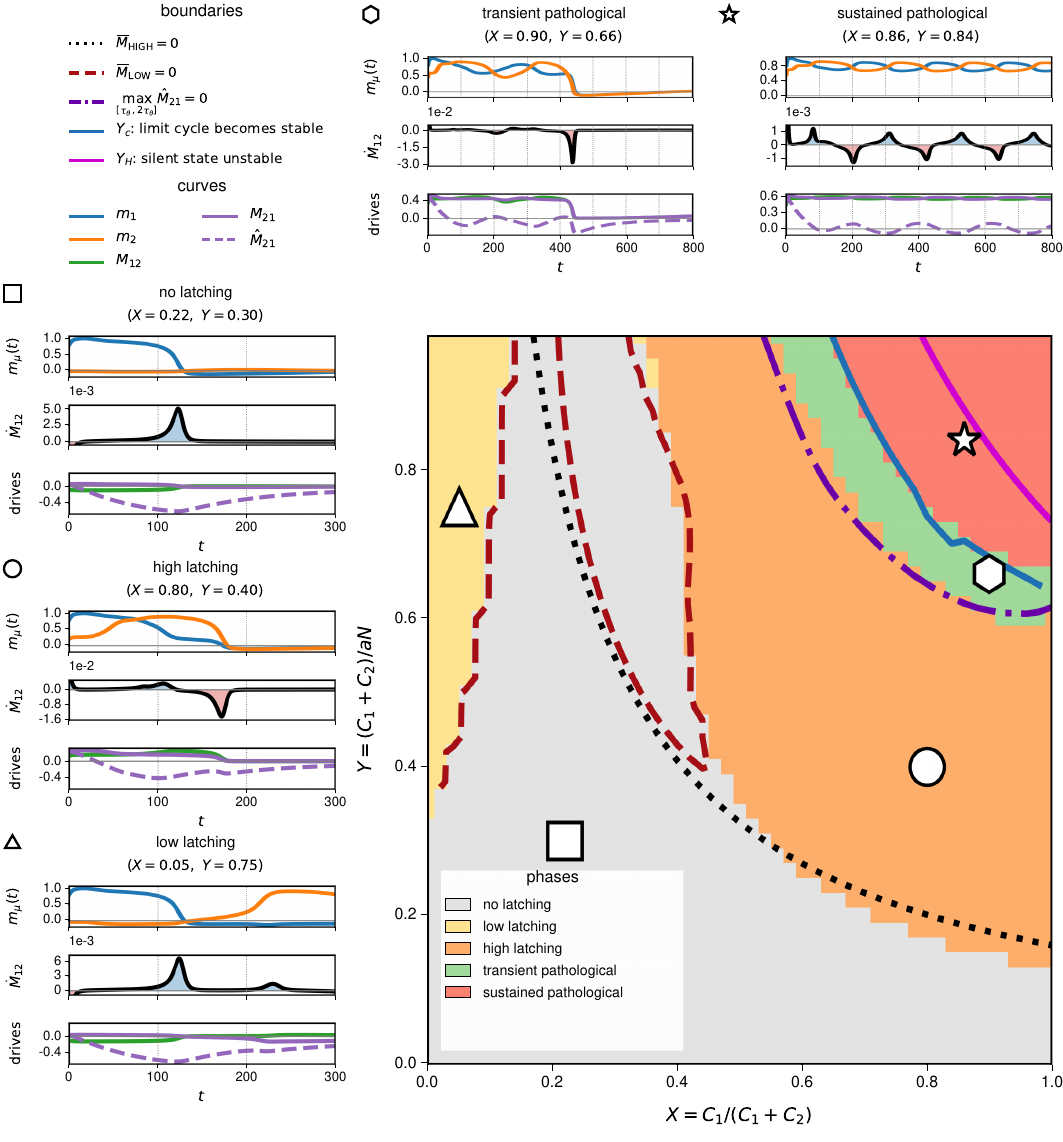}
\caption{
Phase diagram of the latching dynamics as a function of
$X=C_1/(C_1+C_2)$ and $Y=(C_1+C_2)/(aN)$.
The coloured regions identify five dynamical regimes obtained through
finite-size simulations: no latching, low latching, high latching,
transient pathological and sustained pathological latching
(see Appendix~\ref{app:numerical_details_phase} for numerical details).
The superimposed curves mark the loci where the transition observables
change sign: black dotted, $\overline{M}_{\mathrm{HIGH}}=0$;
red dashed, $\overline{M}_{\mathrm{LOW}}=0$; purple dash-dotted,
$\max_{[\tau_\theta,2\tau_\theta]}\hat M_{21}=0$, with
$\hat M_{21}=M_{21}-\theta^x_{xs}$.
The two solid lines follow from the stability analysis of the limit
cycle and the silent fixed point: $Y_c$ (blue), above which the limit
cycle is stable, and $Y_H$ (magenta), above which the silent state
is unstable; between them the network is bistable.
Representative trajectories surround the phase diagram.
The symbol at the upper-left corner of each trajectory panel identifies
the corresponding point in the phase diagram.
Within each trajectory panel, the upper plot shows the overlaps
$m_1(t)$ and $m_2(t)$ with the two condensed patterns;
the middle plot shows $dM_{12}(t)/dt$, with blue and red shading
for its positive and negative parts; and the lower plot shows
the drives $M_{12}$, $M_{21}$ and $\hat M_{21}$.
Parameters: $N=1000$, $p=2$, $S=3$, $a=0.25$, $U=0.3$,
$w=0.8$, $\beta=5$, $\tau_r=10$, $\tau_\theta=100$,
$1/\tau_0=0$.
}
\label{fig:zero_cross_M12}
\end{figure*}

\subsubsection{Diagram of latching transitions}

The resulting phase diagram, together with the representative dynamics inside each phase, are shown in Fig.~\ref{fig:zero_cross_M12}. The central panel reports the phase diagram in the plane $(X,Y)$ (see Sec. \ref{mainmod} for the definitions). 
To illustrate the nature of the different regimes, Fig.~\ref{fig:zero_cross_M12} also shows the time
evolution of the overlaps and of the transition-drive in the five distinct phases. 
For every representative point, the
upper panel shows the overlaps with the two condensed patterns, the middle
panel the derivative $dM_{12}(t)/dt$, and the lower panel the drives $M_{12}$,
$M_{21}$ and $\hat{M}_{21}$. In the middle panels, the intervals in which
$dM_{12}(t)/dt>0$ are shaded in light blue, whereas those in which
$dM_{12}(t)/dt<0$ are shaded in light red.

The curves superimposed on the phase diagram mark the loci where the
observables change sign: $\overline{M}_{\mathrm{HIGH}}$ (black dotted),
$\overline{M}_{\mathrm{LOW}}$ (red dashed) and
$\max_{[\tau_{\theta},2\tau_{\theta}]}\hat{M}_{21}$ (purple dash-dotted),
together with the two lines obtained from the stability analysis, $Y_c$ (solid
blue), where the limit cycle becomes stable, and $Y_H$ (solid magenta), where
the silent state loses its stability.

At low values of $Y$, the phase boundary is primarily controlled by the sign of $\overline{M}_{\mathrm{HIGH}}$. In this regime, the emergence of high-latching is determined by whether the effective transition drive remains positive during the destabilization of the first memory.
For larger values of $Y$, however, the dynamics becomes more complex.
In the upper-left region of the phase diagram, the system displays low-latching: the transition toward the second memory occurs only after the overlap with the first memory has substantially decayed. The boundary of this regime is captured by $\overline{M}_{\mathrm{LOW}}$. Indeed, when $\overline{M}_{\mathrm{LOW}}$ changes sign from positive to negative, the second increase of the effective
drive is absent, and low-latching is no longer observed.

Moving from left to right in the high-$Y$ region of the phase diagram, the system first enters a non-latching phase, where both $\overline{M}_{\mathrm{HIGH}}$ and $\overline{M}_{\mathrm{LOW}}$ are negative. In this regime, the effective drive is not strong enough to induce a transition toward the second memory. By further increasing $X$, one then reaches an intermediate regime in which both $\overline{M}_{\mathrm{HIGH}}$ and $\overline{M}_{\mathrm{LOW}}$ becomes positive, but the overlap with the second memory still does not reach large values after the transition. 
Interestingly, the final boundary between no-latching and high-latching is controlled by the change of sign of $\overline{M}_{\mathrm{LOW}}$. The dynamical origin of this additional constraint remains an open question and will be addressed in future work.

Increasing $X$ still further in the high-$Y$ region, the system first enters
the transient pathological regime, whose boundary is signalled by the change of sign of the maximum of $\hat{M}_{21}$ (purple dash-dotted line); it then
becomes bistable (blue solid line), so that it can end up either on the limit
cycle or on the silent state depending on the initial condition; and finally
the silent state loses its stability (magenta solid line), after which the only possible outcome is sustained pathological latching.

\subsection{Suppression of latching in the binary case}

The binary case $S=1$ provides a minimal setting to understand whether adaptation alone is sufficient to sustain memory transitions.
In this case, the two-condensed {\it ansatz} contains only 
four pattern configurations $g \in \{xx,xs,sy,ss\}$, since the $xy$ site class 
is absent for $S=1$.
Consequently, the overlap structure between the two 
memories is controlled only by the single parameter $c_{xx} = \frac{C_1}{N}$.

The effective transition drive in the binary case assumes the form
\begin{eqnarray}
M_{12}(\{\sigma_g \})
\simeq
\frac{C_1 (1-2a)}{Na}  \sigma_{xx}^x
-
2a \Big( 1- \frac{C_1 }{Na} \Big) \sigma_{xs}^x.
\end{eqnarray}
When the network state is close to the first memory, the active units satisfy 
approximately $\sigma_{xx}^x \simeq \sigma_{xs}^x \simeq 1$. Substituting this
approximation into the above equation and imposing $M_{12}(\{\sigma_g \}) > 0$ on the effective drive, we obtain the latching condition
\begin{equation}
\frac{C_1}{Na} > 2 a . 
\label{utrr}
\end{equation}
For two memories drawn independently from the random pattern distribution, where each unit is active with probability \(a\), the typical value of their overlap is $ C_1 \simeq Na^2$ .  Therefore, the condition \(C_1>2Na^2\) is not typically satisfied.

\begin{figure}[t]
\centering
\includegraphics[width=\linewidth]{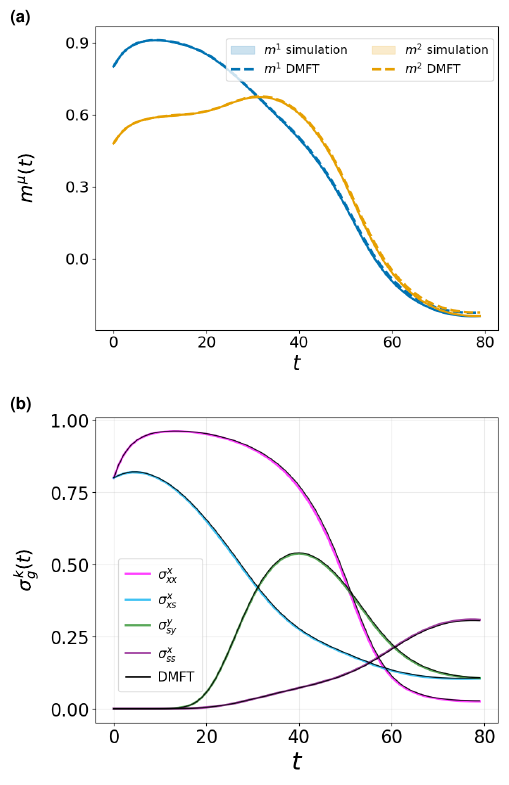}

\caption{
Finite-size simulations in the binary case \(S=1\).
Parameters: \(N=1000\), \(p=2\) (\(\alpha=0\)),
\(a=0.25\), \(U=0.3\), \(\beta=5\), \(\Delta t=1\),
\(\tau_r=10\), \(\tau_\theta=25\), \(w=0.8\),
\(C_1/(Na)=0.7\).
Top: overlaps \(m^1(t)\) and \(m^2(t)\).
Bottom: representative activities \(\sigma_g^k\).
}
\label{fig: S1_latch}
\end{figure}

The situation is different in the multi-state case, as the $xy$ units provide additional transition channels between memories through partially overlapping but differently colored states. In the binary case, these channels are absent. However, weak crossover-like behavior still occurs in the binary case, but only
after adaptation has substantially suppressed the activity of the first memory. Indeed, as
shown in Fig. \ref{fig: S1_latch} through finite-size simulations, the  
activities $\sigma_{xs}^x$ and $\sigma_{sy}^y$ can transiently cross for sufficiently
large $C_1$. However, these transitions remain strongly suppressed compared to the $S>1$ case and do not generate robust high-latching dynamics.

\subsection{Latching transitions for $\alpha>0$}

We now turn to the extensive-load regime, $\alpha>0$, where the non-condensed memories generate both colored noise and retarded self-interactions in the effective DMFT dynamics. In contrast to $\alpha=0$, adaptive transitions are no longer governed only by the geometric structure of the pattern overlaps controlled by $X$ and $Y$. The extensive background of non-condensed patterns dynamically reshapes the effective temporal organization of the single-site dynamics.

This effect is encoded in the time-dependent memory kernel $K(t,t')$. In the effective dynamics, the retarded contribution has the form
\[
h_{\mathrm{mem}}^k(t)
=\frac{1}{\tilde a}
\int_0^t dt'\sum_{l=1}^SC_{kl}\,K(t,t')\,\sigma^l(t') ,
\]
so that the present field depends on the past activity through $K(t,t')$. The kernel can therefore be interpreted as an additional dynamical memory generated by the non-condensed patterns. During latching transitions, this effective memory develops a clear dynamical signature.

\begin{figure*}[t]
\centering
\includegraphics[width=\linewidth]{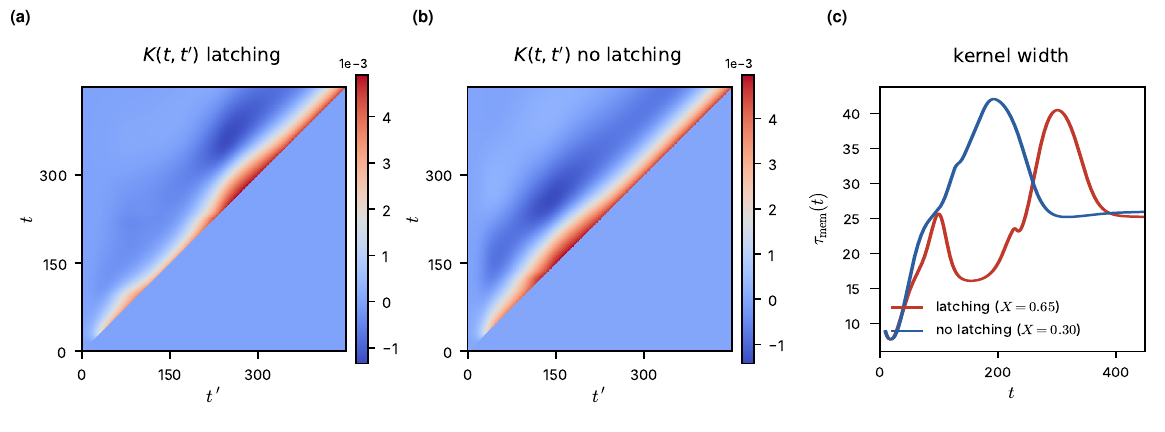}

\caption{
Memory kernels for two representative memory pairs at extensive load
($\alpha=0.1$). (a) latching trajectory with \(X=0.65\), \(Y=0.8\),
lying inside the high-latching region shown in
Fig.~\ref{fig:alpha_diagram}. The kernel width develops a peak around
\(t\sim\tau_\theta=10\tau_r=100\).
(b) non-latching trajectory with \(X=0.45\), \(Y=0.8\),
where the kernel width increases progressively in time before decaying
after the transition to the silent state.
(c) Kernel width \(\tau_{\rm mem}(t)\), defined as the time lag at which
\(K(t,t')\) falls to half of its maximum over \(t'\le t\), for the two
trajectories: two peaks in the latching case (red) and a single peak in the
non-latching case (blue). 
Other parameters are as in Fig.~\ref{fig:zero_cross_M12}.
}
\label{fig:mem_kernel_latch_no_latch}
\end{figure*}

Figure~\ref{fig:mem_kernel_latch_no_latch} shows the memory kernel for two representative pairs of memories. In the latching case, the width of the kernel develops a pronounced peak around the adaptation timescale and a second peak close to the silent state, whereas in the non-latching case the kernel broadens progressively in time and peaks only when the network reaches the silent state. 

\subsubsection{Effective memory timescale}
\begin{figure*}[!tp]
\centering
\includegraphics[
  width=\textwidth,
  height=0.75\textheight,
  keepaspectratio
]{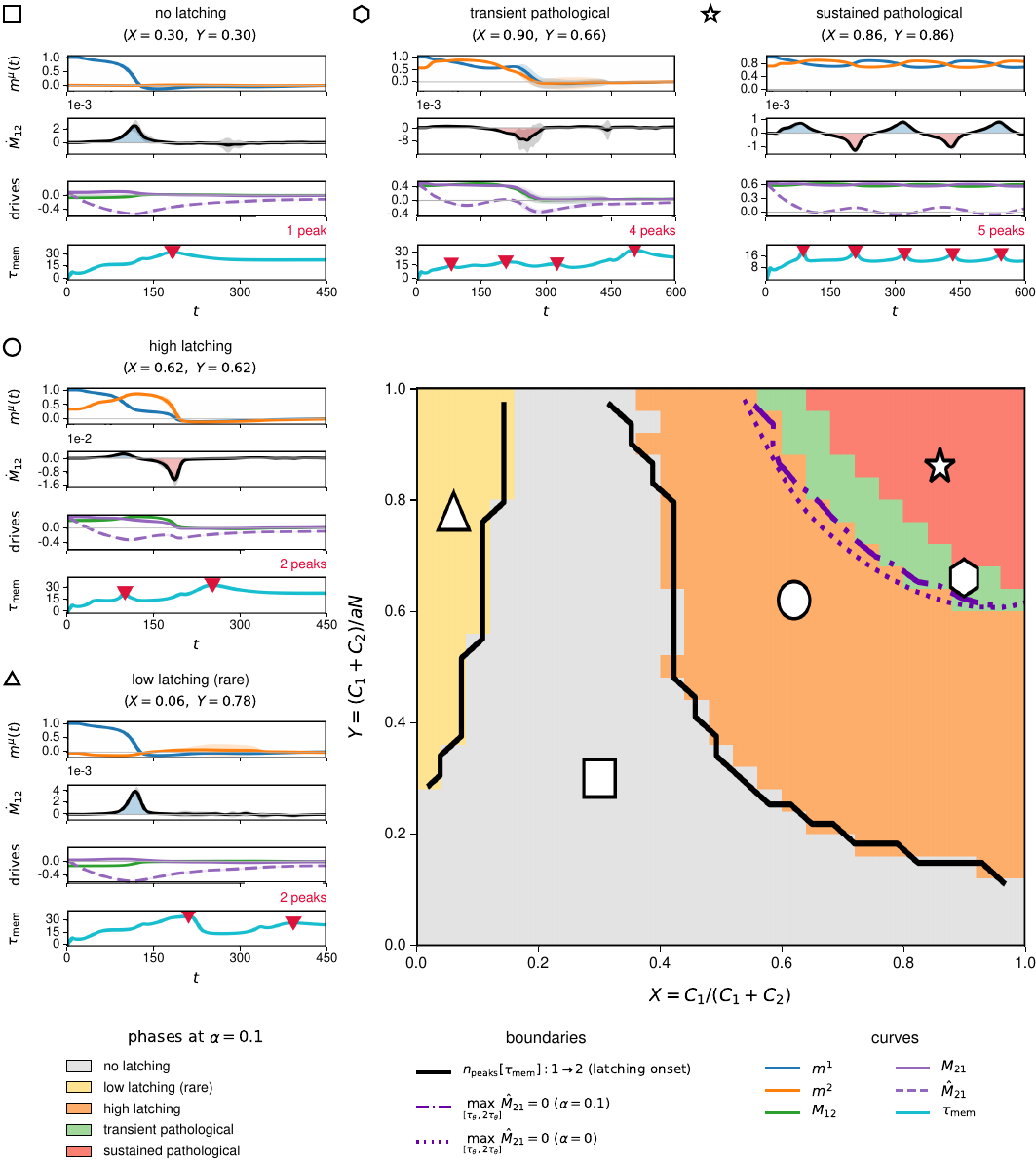}
\caption{
Phase diagram of the latching dynamics at $\alpha=0.1$,
in the same format as Fig.~\ref{fig:zero_cross_M12}.
The symbol at the upper-left corner of each trajectory panel identifies
the corresponding point in the phase diagram.
The trajectory panels include a fourth plot showing the half-width
$\tau_{\mathrm{mem}}(t)$ of the memory kernel, whose peaks are
marked by red triangles.
The black solid line marks the locus where
$n_{\mathrm{peaks}}[\tau_{\mathrm{mem}}]$ goes from $1$ to $2$,
i.e. where the first transition appears.
The purple dash-dotted and dotted lines show
$\max_{[\tau_\theta,2\tau_\theta]}\hat M_{21}=0$
at $\alpha=0.1$ and $\alpha=0$, respectively; they nearly coincide.
The yellow region contains points where a low-latching event
is still observed, with a mean probability of $5\%$.
Curves are averages over $16$ realisations of the disorder
at $N=800$, with bands of $\pm1$ standard deviation.
The two pathological examples are shown up to $t=600$,
and the other three up to $t=450$.
Other parameters are as in Fig.~\ref{fig:zero_cross_M12},
with $N=800$ and $\alpha=0.1$.
}
\label{fig:alpha_diagram}
\end{figure*}

To quantify this effect, we introduce an effective memory timescale $\tau_{\mathrm{mem}}(t)$ that characterizes the temporal width of the kernel. For a fixed observation time $t$, the kernel $K(t,t')$ reaches a maximum near $t'=t$ and decays as one moves backward in time. We define $\tau_{\mathrm{mem}}(t)$ as the half-width of the kernel.
More explicitly, if
\begin{equation}
K_{\max}(t)=\max_{0 \leq t^{\prime} \leq t}K(t,t'), \nonumber
\end{equation}
then $\tau_{\mathrm{mem}}(t)$ is defined by
\begin{equation}
K\bigl(t,t-\tau_{\mathrm{mem}}(t)\bigr)
=
\frac{1}{2}K_{\max}(t). \nonumber
\end{equation}
The behavior of $\tau_{\mathrm{mem}}(t)$ provides a clear distinction between non-latching, high-latching, and pathological-latching trajectories. In all three regimes, $\tau_{\mathrm{mem}}(t)$ exhibits a final peak when the network decays into the silent state. In non-latching trajectories, this is the only prominent peak, with $\tau_{\mathrm{mem}}(t)$ growing approximately linearly over the relevant time window before the final decay. In the high-latching and low-latching regime, an additional peak appears on the adaptation timescale, associated with the transition from one memory to another. In the pathological-latching regime, instead, $\tau_{\mathrm{mem}}(t)$ displays a sequence of peaks, occurring each time the activities of the two memories cross during their repeated alternation.
Thus, the number of peaks of the memory timescale $\tau_{\mathrm{mem}}(t)$ can be used as a dynamical probe of the latching regime at extensive load, as one can see in fig \ref{fig:alpha_diagram}. 

The physical reason is implicit in the definition of the response. 
\begin{equation}
\tilde G(t,t')=\frac{1}{
\tilde{a}
}\sum_g c_g\sum_{kl}C_{kl}\,
\left.\frac{\delta\sigma^k_g(t)}{\delta h^l_g(t')}\right. 
\end{equation}
and at equal times the
response to a field $h_g^l(t)$, applied to $r_g^l(t)$, is the Jacobian of the softmax,
\begin{equation}
\begin{aligned}
\frac{\delta\sigma_g^k(t)}{\delta h_g^l(t)}
= \sum_{m,g'}
\frac{\delta\sigma_g^k}{\delta r_{g'}^m}
\frac{\delta r_{g'}^m}{\delta h_g^l}
&= \frac{1}{\tau_r}A_g^{kl}(t),\\
A_g^{kl}
&= \beta\,\sigma_g^k
\bigl(\delta_{kl}-\sigma_g^l\bigr).
\end{aligned}
\end{equation}
so that
\begin{equation}
\tilde G(t,t)=\frac{1}{\tau_r\,
\tilde{a}
}\sum_g c_g\,\mathrm{Tr}\!\left[C\,A_g(t)\right].
\end{equation}
Using $C_{kl}=(a/S)\delta_{kl}-(a/S)^2$ and writing $Q_g=\sum_k\sigma^k_g$ for the
total activity of class $g$, the trace can be computed explicitly,
\begin{equation}
\mathrm{Tr}\!\left[C A_g\right] = \beta\Biggl[
\frac{a}{S}\sum_k \sigma_g^k
\bigl(1-\sigma_g^k\bigr) - \left(\frac{a}{S}\right)^2
Q_g\bigl(1-Q_g\bigr)
\Biggr].
\end{equation}
The first term dominates, while the second is smaller by an additional factor of (a/S). The first term $\sum_k\sigma^k_g(1-\sigma^k_g)$ vanishes if and only if every $\sigma^k_g\in\{0,1\}$, that is when one
memory is fully retrieved ($Q_g=1$) or in the silent state ($Q_g=0$).
Neglecting the contribution coming from the activity of units in a color different from a condensed one (that is usually smaller than the activity in the condensed color as one can see in Fig. \ref{fig: overlap alpha=0.1}), the contribution of the first term to the equal time response function is proportional to:
\begin{equation}
\label{eq: resp_contributions}
\begin{aligned}
&c_{xy}\Bigl[
\sigma_{xy}^x(1-\sigma_{xy}^x)
+\sigma_{xy}^y(1-\sigma_{xy}^y)
\Bigr] + c_{xx}\sigma_{xx}^x(1-\sigma_{xx}^x)\\
&\quad + c_{sy}\sigma_{sy}^y(1-\sigma_{sy}^y) + c_{xs}\sigma_{xs}^x(1-\sigma_{xs}^x).
\end{aligned}
\end{equation}
Each contribution is small when the network is close to a condensed memory or to the silent state. During a transition from the first to the second memory, the $xy$ units redistribute their activation between colors $x$ and $y$. Consequently, the term proportional to $c_{xy}$ in expression \eqref{eq: resp_contributions} reaches its maximum value of $1/2$ when $\sigma_{xy}^x \simeq \sigma_{xy}^y \simeq 1/2$, typically near the crossing of the two overlaps. Similarly, the $sy$ and $xs$ units contribute at most $1/4$ each, with their contributions reaching these maxima when the corresponding units are equally distributed between the silent and active states.
The units that are silent in both
memories, which have all $S$ colors weakly and equally active, add an
approximately constant background to the response.

The reason why $\tau_{\mathrm{mem}}(t)$ develops peaks in correspondence with
the transitions is that it is controlled by the equal-time response
$\chi(t)\equiv\tilde G(t,t)$. Assuming that the response function has 
a peak at equal time and decays exponentially with a time scale $\tau_G$, then we can write
$\tilde G(t,t')=\chi(t)\,e^{-(t-t')/\tau_G}$, and the series
$K(t,t')=\alpha\sum_{n\ge1}\tilde G^{\,n}(t,t')$ can be summed exactly (see
Appendix~\ref{app: mem_kernel_timescale}):
\begin{equation}
K(t,t')=\alpha\,\chi(t)
\exp\Big[-\frac{1}{\tau_G}\int_{t'}^{t}du\,\big(1-\tau_G\chi(u)\big)\Big].
\label{eq:K_chi}
\end{equation}
For $\chi<1$ the kernel is maximal at $t'=t$, and the definition of
$\tau_{\mathrm{mem}}$ gives
\begin{equation}
\int_{t-\tau_{\mathrm{mem}}(t)}^{t}du\,\big(1-\tau_G\chi(u)\big)=\tau_G\ln 2 .
\label{eq:taumem_chi}
\end{equation}

When $\chi$ varies little over the $\tau_{mem}$ window:
\begin{equation}
    \tau_{\mathrm{mem}}(t) \simeq \frac{\tau_G\ln2}{1-\tau_G\chi(t)}
\end{equation}
Hence
$\tau_{\mathrm{mem}}(t)$, for $\chi<1$, is an increasing function of the susceptibility. See appendix~\ref{app: mem_kernel_timescale} for a complete derivation and numerical validation of the formula in Eq. \ref{eq:taumem_chi}.

As shown in Fig.~\ref{fig:alpha_diagram}, in the non-latching regime $\tau_{\mathrm{mem}}$ exhibits a single peak, associated with the final decay into the silent state. In both the low- and high-latching regimes, two peaks are observed: one when the two memory overlaps cross during the transition, and one at extinction. In the pathological-latching regime, instead, $\tau_{\mathrm{mem}}$ peaks at each crossing of the two overlaps as the network repeatedly alternates between the same two memories. Thus, counting the peaks of $\tau_{\mathrm{mem}}$ is equivalent to counting the memory switches, with one additional peak corresponding to extinction:
\begin{equation}
n_{\mathrm{peaks}}[\tau_{\mathrm{mem}}]=n_{\mathrm{switch}}+1,
\nonumber
\end{equation}
The peaks can be
located, in practice, with a standard prominence-based algorithm (\texttt{scipy.signal.find\_peaks}).

\subsubsection{Suppression of low latching at finite load.}
\begin{figure}[t]
\centering
\includegraphics[width=\linewidth]{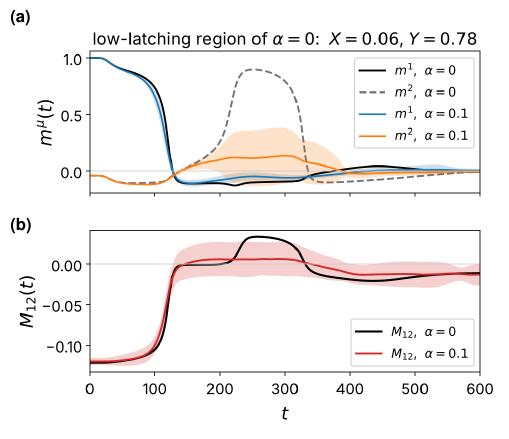}

\caption{
Suppression of low latching at finite load for $X=0.06$, $Y=0.78$.
Top: overlaps with the two condensed patterns at $\alpha=0$
(black and grey dashed lines) and at $\alpha=0.1$
(blue and orange; mean over 24 disorder realisations,
with bands of $\pm1$ standard deviation).
Bottom: transition drive $M_{12}(t)$. At $\alpha=0.1$,
its late peak is strongly reduced, suppressing low latching.
Other parameters as in Fig.~\ref{fig:zero_cross_M12}, with $N=800$.
}
\label{fig:low_latching}
\end{figure}

Low latching is the regime most strongly suppressed by the load. We illustrate this at $X=0.06$, $Y=0.78$, a
point that lies deep inside the low-latching region of the $\alpha=0$ diagram,
simulating $24$ independent realisations of the disorder at $N=800$
(Fig.~\ref{fig:low_latching}).
Averaged over all realisations the peak overlap falls from $0.90$
to $0.22$, i.e. the typical trajectory no longer
retrieves the second memory at all.

The origin of this effect is visible in the drive. The corrected transition
drive $M_{12}(t)$, which at $\alpha=0$ develops a second, delayed rise after the
first memory has decayed, is reduced at $\alpha=0.1$:
its reduction is enough to suppress low-latching trajectories. High latching, whose drive is an order of magnitude larger, is unaffected. 

\section{Discussion}

In this work, we have used dynamical mean-field theory to study latching dynamics in Potts associative networks. The central question was whether slow adaptation, expressed as adapting thresholds \cite{horn1989neural}, can generate on its own non-random transitions even in standard networks, in which units represent individual neurons. Simulations had indicated that while specific transitions can be associatively learned in the connectivity matrix of a standard network \cite{herrmann1993neural}, spontaneous ones require additional representational structure, i.e., the units in the network should be able to access distinct activity states, e.g. modelling distinct attractors in a local patch of cortex \cite{Treves2005}. The Potts network offers a convenient formulation of this issue, as the two cases are expressed simply by setting $S=1$ or $S>1$. Our analysis in the
finite-load regime, $\alpha = 0$, confirms and explains the insight from the simulations. We show that transitions between memories are controlled by the overlap between them, through an effective drive $M_{12}$. In the binary case, $S=1$, adaptation can destabilize the currently retrieved attractor but cannot efficiently generate a directed transition toward another memory while the first memory remains active. At finite load $\alpha>0$, non-condensed memories generate retarded self-interactions and colored fluctuations, substantially suppressing the low-latching regime. The DMFT memory kernel develops a characteristic temporal structure associated with transitions between memories, but the basic result essentially holds. Thus, spontaneous, non-random transitions require three ingredients: destabilization of activated memory through adaptation, a suitable overlap between memories (that is not necessarily typical of random independent memories), but more fundamentally also local categorical degrees of freedom that provide directed transition channels. This adds to other respects in which the $S=1$ and $S>1$ networks significantly differ, eg in their retrieval properties \cite{ryu2026retrieval,kim2026glassy}.

In terms of implementing associative, content-addressable memory, Potts networks can represent interacting cortical patches endowed each with several local attractor states, while long-range connections bind local states into global memory patterns \cite{o1992short}. Latching dynamics adds a potential extra functionality \cite{Treves2005} which, our results confirm, is not available in a standard binary neural net, as it requires multiple local states to open up {\em transition channels} between memories. In particular, the $xy$ pattern configuration, which exists only for $S>1$, includes units that transition from their active state in one memory to their active state in the other. Through such a route, which amounts to `swiveling' around the units active in the same state in both memories, adaptation can redirect the network from one global attractor to another before the first one has fully decayed.

The phase diagram at $\alpha=0$ clarifies how the two components of memory similarity constrain possible transitions. The variables $X$ and $Y$ separate two aspects of memory overlap: the fraction of jointly active units sharing the same categorical state and the total fraction of common units active in both memories. The effective drive $M_{12}$ shows that both quantities determine the access to a second memory during adaptation. In this sense, the overlap between memories defines a transition geometry: some memory pairs are dynamically reachable under adaptation, while others are not. This provides a mechanistic interpretation of latching as motion through a structured memory space rather than as random hopping between attractors of independent memories.

More broadly, the latching trajectories observed here belong to a class of metastable neural dynamics which model cognitive processes characterized as a sequence of transiently stable states, or chaotic itinerancy among {\em attractor ruins} in the language of Tsuda \cite{itoh1997chaotic,tsuda2001plausibility,kaneko2002coupled}, rather than those that reduce to convergence onto a single fixed point \cite{Rabinovich2008Science,Rabinovich2011}. The network remains close to a retrieved attractor---a \textit{ruin}---for a certain dwell time, e.g. in the order of a hundred {\em msec}, then adaptation gradually reduces its stability, until it collapses onto another ruin depending on their overlap, and so on, indefinitely \cite{RussoTreves2012}. Importantly, however, in the case of latching the metastable dynamics are spontaneous and not instructed: instructing transitions does not work too well with Potts networks \cite{kang2019challenge,mun2026latching}.

In the extensive-load regime, non-condensed memories generate both colored noise
and a retarded self-interaction kernel \cite{Coolen2001}. Its effective memory
timescale $\tau_{\mathrm{mem}}(t)$  displays at least one peak, at the extinction
of the activity, and each latching transition contributes one further peak,
located at the crossing of the two overlaps. The number of peaks
can be used to identify the latching region directly from the DMFT kernel. The peaks
signal the passage through configurations
of maximal susceptibility: the equal-time response is the Jacobian of the softmax,
which vanishes both when a memory is fully retrieved and in the silent state, and
is largest when the activity is shared between two states.

This kernel-based signature also places the present work within the broader statistical physics program of reducing high-dimensional recurrent dynamics to a description in terms of low-dimensional self-consistent order parameters. Classical dynamical mean-field approaches have shown how disorder can generate effective noise, memory kernels, and complex temporal regimes in recurrent networks \cite{Sompolinsky1988,Coolen2001}. Recent work has emphasized how low-dimensional structure embedded in random recurrent connectivity can shape population dynamics and computations \cite{MastrogiuseppeOstojic2018}. In our setting the low dimensionality is ever transient: at each transition, the `condensed' memories, i.e., the pair undergoing the transition, provide structured directions in configuration space, while the non-condensed memories generate a disordered dynamical background. The resulting latching dynamics is therefore shaped by the interplay between transient low-dimensional memory geometry and quenched disorder.

Several limitations remain. First, the present analysis has focused on transitions between two condensed memories. Although the effective single-site dynamics, Eq. (\ref{eq: single site eq}), holds for an arbitrary number of condensed patterns, to consider protracted autonomous latching sequences involving many memories one needs to enumerate all pattern configurations
and study how the system selects its states among multiple possible attractors. Second, the stored patterns considered here are random, apart from controlled pairwise overlap parameters. Semantic memories in the cortex are expected to have richer higher-order correlations, which may strongly affect both transition probabilities and DMFT kernels. Third, adaptation is modeled here as a slow threshold variable. A wealth of other mechanisms can be included, such as synaptic depression, or multiple forms of slow and fast inhibition \cite{tremblay2016gabaergic}, which would modify the conditions for latching \cite{RussoTreves2012}. Finally, although the memory-kernel timescale provides a useful dynamical probe of latching at finite load, a more complete analytical theory of the different regimes remains to be developed.

Even within the random memory assumption, it would be useful to derive a full phase diagram that e.g. shows how latching dynamics depends on the connectivity and on the (possibly heterogeneous \cite{ryom2023speed}) multiplicity of local attractor states in a Potts network. This may corroborate or falsify the hypothesis that latching dynamics emerge through a phase transition in the evolution of the cortex of certain mammals, following quantitative increases in its local and long-range connectivity \cite{Treves2005}. This is left for future work.  
\section{Conclusion}

We have shown that slow adaptation alone is insufficient to generate robust non-random transitions between memories. Instead, latching requires suitable overlap together with the local categorical structure modelled by multi-state Potts units. In the binary case, adaptation destabilizes retrieved memories but fails to efficiently redirect the dynamics toward alternative attractors. At finite load, latching transitions are additionally associated with a restructuring of the DMFT memory kernel and its effective timescale. These results suggest that multi-state representations may play a fundamental dynamical role in enabling autonomous memory-to-memory transitions in recurrent cortical systems.
\begin{acknowledgments}
\end{acknowledgments}

Preliminary stages of this study saw the participation of Dr Kwang Il Ryom, whose contribution is gratefully acknowledged. F.L.M. acknowledges financial support from CNPq (Grants No 402487/2023-0 and No 310255/2025-2) and the ICTP through the Associates Programme (2023-2028).

\bibliography{references_all}

\clearpage
\onecolumngrid
\startsupplementnum
\appendix

\section{Dynamical mean-field theory}\label{app: dis_average} 

In this appendix, we derive the effective single-site equation for $N \rightarrow \infty$ using dynamical mean-field theory (DMFT). We consider that \(B\) condensed patterns are kept fixed and we average over the remaining \(p-B\) patterns, with $p = \alpha N$. 
We focus on the case \(B=1\) for simplicity. The extension to an arbitrary number of condensed patterns, \(B>1\), is straightforward.

\subsection{The disorder average and saddle-point integral}

The goal is to compute the disorder average $\mathbb{E}_\xi\big[ \dots \big]$
over the non-condensed patterns
\begin{equation}
\overline{\mathcal{Z}[\boldsymbol{\psi}]}
=
\mathbb{E}_\xi\big[\mathcal{Z}[\boldsymbol{\psi}]\big], 
\end{equation}
where $\mathcal{Z}[\boldsymbol{\psi}]$ is the generating functional of the
probability $\mathcal{P}\left[\boldsymbol{r}\right]_{t\in[0,T]}$ of observing a dynamical path in $t \in [0,T]$,
\begin{equation}
    \quad
\mathcal{Z}[\boldsymbol{\psi}]
=
\Bigg\langle
\exp\Bigg(
\int_{0}^{T}\! dt \sum\limits_{i=1}^N \sum\limits_{k=1}^S
\psi_i^k(t) \sigma_i^k(t)
\Bigg)
\Bigg\rangle.
\end{equation}
The brackets $\big\langle \dots \big\rangle$ denote the average 
with the path-probability
\begin{align}
\mathcal{P}\left[\boldsymbol{r}\right]_{t\in[0,T]}
=
\prod_{i=1}^N\prod_{k=1}^S \delta_{(F)}
\big[ \tau_r \dot r_i^k(t) - h_i^k(t) + r_i^k(t) + \Omega_i^k(t)  \big]
\end{align}
of the fields $\{ r_{i}^{k}(t) \}$ ($i=1,\dots,N$ and $k=1,\dots,S$). In the above
expression, the (external) input field is given by
\begin{equation}
h_i^k(t) = \sum_{j=1}^N \sum_{l=1}^S J_{ij}^{kl} \, \sigma_j^l(t), 
\end{equation}
while $\Omega_{i}^{k}(t)$ reads
\begin{equation}
\Omega_{i}^{k}(t) = \theta_{i}^{k}(t) - 
w \Big( \sigma_i^k(t) - \frac{1}{S} \sum_{l=1}^{S} \sigma_i^{l}(t) \Big).
\label{rewd}
\end{equation}
Writing the $\delta$-functional as a Fourier transform, we can introduce the conjugate fields
$\hat r_i^k(t)$ 
\begin{equation}
\mathcal{P}\left[\boldsymbol{r}\right]_{t\in[0,T]}
=
\int D\boldsymbol{\hat r}\; 
\exp\Bigg( i
\int_0^T dt \sum\limits_{i,k}
\hat r_i^k(t)
\big(\tau_r \dot r_i^k(t) - h_i^k(t) + \Omega_i^k(t) +r_i^k(t)\big),
\Bigg)
\end{equation}
with $D\boldsymbol{\hat{r}} = \prod_{i=1}^N \prod_{k=1}^S D \hat{r}_{i}^{k}(t)$.
Thus, the disorder average of the generating functional assumes the convenient form:
\begin{align}
\overline{\mathcal{Z}[\boldsymbol{\psi}]}
=
\int D\boldsymbol{r}\,D\boldsymbol{\hat r}\;
&
\exp\Bigg(
\int_0^T dt\;\sum\limits_{i=1}^N \sum\limits_{k=1}^S \big[ \psi_i^k(t)\sigma_i^k(t) +  i\hat r_i^k(t) \big(\tau_r \dot r_i^k(t) + \Omega_i^k(t) + r_i^k(t) 
\big) \big]
\Bigg)
\nonumber \\
& \times
\mathbb{E}_\xi \Bigg[ 
\exp\Bigg(
-i \int_0^T dt \sum\limits_{ij=1}^N \sum\limits_{kl=1}^S \hat r_i^k(t) J_{ij}^{kl} \sigma_j^l(t)
\Bigg) 
\Bigg]. 
\label{eq: GF}
\end{align}

We are now in a position to explicitly compute the disorder average $\mathbb{E}_\xi[\dots]$ over the non-condensed patterns $p > 1$. Throughout the rest of this section and in order to simplify the notation, we will omit the summation limits of the sums over colors and sites. The sums over sites \((i,j)\) run from \(1\) to \(N\), while the sums over colors \((k,l)\) run from \(1\) to \(S\). 

Let us substitute the Hebbian interactions, Eq. (\ref{eq:J}), and define 
\begin{equation}
\mathcal{F} = 
\mathbb{E}_\xi \Bigg[ 
\exp\Bigg(
-\frac{i}{N_a}
\int_0^T dt\,
\sum\limits_{i,j(\ne i)}
\sum\limits_{k,l}
\sum\limits_{\mu=1}^p
\hat r_i^k 
\eta_{ik}^\mu \eta_{jl}^\mu
\sigma_j^l
\Bigg)  
\Bigg],
\end{equation}
where
\begin{equation}
    \eta_{ik}^\mu =\delta_{\xi_i^\mu, k} - a/S \quad (k=1,\dots,S)
    \label{hui}
\end{equation}
and $N_a = Na(1-a/S)$. We can now single out the first pattern from the sum over $\mu$ and write 
\begin{align}
\mathcal{F} \simeq &\;\exp\Bigg({
    - i \int_0^T dt \frac1{N_a} 
    \left[ \sum_{i,k} \hat r_i^k(t) \eta_{ik}^1\right]
    \left[ \sum_{i,k} \sigma_i^k(t) \eta_{ik}^1\right]
    }\Bigg)\nonumber\\[3mm]
    \times 
    \mathbb{E}_\xi \Bigg[   
    &\exp\Bigg({
    - i \sum_{\mu > 1} \int_0^T dt 
    \left[ \sum_{i,k} \frac{\hat r_i^k(t) \eta_{ik}^\mu}{\sqrt{N_a}}\right]
    \left[ \sum_{i,k} \frac{\sigma_i^k(t) \eta_{ik}^\mu}{\sqrt{N_a}}\right]
    +
    \sum_{\mu > 1} \int_0^T dt \frac{i}{N_a} \sum_i
    \left[ \sum_{k} \hat r_i^k(t) \eta_{ik}^\mu\right]
    \left[ \sum_{k} \sigma_i^k(t) \eta_{ik}^\mu\right]
    }\Bigg)
    \Bigg].
    \label{yu}
\end{align}
We have neglected the terms of $\mathcal{O}(N^0)$ in the exponent
of the above expression, since they do not contribute to the generating functional for $N \rightarrow \infty$. To decouple the non-condensed patterns and factorize the expression inside the average into a product over patterns, we introduce
appropriate $\delta$-functionals and their corresponding Fourier representations, yielding

\begin{align}
\mathcal{F} \simeq \;&\exp\Bigg({
    - i \int_0^T dt \frac1{N_a} 
    \left[ \sum_{i,k} \hat r_i^k(t) \eta_{ik}^1\right]
    \left[ \sum_{i,k} \sigma_i^k(t) \eta_{ik}^1\right]
    }\Bigg)\nonumber \nonumber\\[3mm]  
    & \times \prod_{\mu=2}^p 
    \int Dx_{\mu} Dy_{\mu} D\hat{x}_{\mu} D\hat{y}_{\mu}
    \;\;
    \exp\Bigg({i\int_0^Tdt \; 
    \Big[ 
    -x_\mu(t)y_\mu(t)
    + \hat x_\mu(t)x_\mu(t)
    + \hat y_\mu(t)y_\mu(t)
    \Big] }\Bigg) \nonumber\\
    &\times    
    \prod_{i=1}^N  \mathbb{E}_\xi \Bigg[   
    \exp{\Bigg(
    -\frac{i}{\sqrt{N_a}}\int_0^Tdt \sum_{k}
    \hat x_\mu(t)\hat r_i^k(t)\eta_{ik}^\mu
    -\frac{i}{\sqrt{N_a}}\int_0^Tdt \sum_{k}
    \hat y_\mu(t) \sigma_i^k(t)\eta_{ik}^\mu
    \Bigg)}
    \nonumber\\
    &\times
    \exp{\Bigg(
    \frac{i}{N_a} \sum_{kl}  \eta_{ik}^\mu \eta_{il}^\mu  \int_0^T dt \,  \hat r_i^k(t)\sigma_i^l(t)
    \Bigg)}
    \Bigg].
\end{align}
For large $N_a$, we expand the argument of $\mathbb{E}_\xi [ \dots ]$ up to $\mathcal{O}(N_{a}^{-1})$, 
\begin{align}
    \mathcal{F} &\simeq \exp\Bigg({
    - i \int_0^T dt \frac1{N_a} 
    \left[ \sum_{i,k} \hat r_i^k(t) \eta_{ik}^1\right]
    \left[ \sum_{i,k} \sigma_i^k(t) \eta_{ik}^1\right]
    }\Bigg)\nonumber\\[3mm]  
    & \times \prod_{\mu=2}^p \prod_{i=1}^N
    \int Dx_{\mu} Dy_{\mu} D\hat{x}_{\mu} D\hat{y}_{\mu} 
    \;\;
    \exp\Bigg({i\int_0^Tdt \; 
    \Big[ 
    -x_\mu(t)y_\mu(t)
    + \hat x_\mu(t)x_\mu(t)
    + \hat y_\mu(t)y_\mu(t)
    \Big] } \Bigg)\nonumber\\
    &\times  
    \mathbb{E}_\xi \Bigg[  
    1
    -\frac{i}{\sqrt{N_a}} \sum_{k} \eta_{ik}^\mu V_{i\mu}^k
    + \frac{i}{N_a} \sum_{k} (\eta_{ik}^\mu)^2  U_i^{kk}
    + \frac{i}{N_a} \sum_{k\ne l} \eta_{ik}^\mu \eta_{il}^\mu U_i^{kl}
    -\frac{1}{2 N_a}\sum_{kl} \eta_{ik}^\mu \eta_{il}^\mu V_{i\mu}^{k} V_{i\mu}^{l}
    \Bigg], \label{ew1}
\end{align}
in which 
\begin{equation}
    V_{i \mu}^{k} = \int_{0}^{T}  dt \big[  \hat{x}_{\mu}(t) \hat{r}_{i}^{k}(t)  + \hat{y}_{\mu}(t) \sigma_{i}^{k}(t)   \big]
\end{equation}
and
\begin{equation}
    U_i^{kl} = \int_{0}^{T}  dt \, \hat{r}_{i}^{k}(t) \sigma_{i}^{l}(t).
\end{equation}
The last line in Eq. (\ref{ew1}) includes only the dominant contributions 
to the generating functional for $N \gg 1$. 
Given that $\mathbb{E}_\xi[ \delta_{\xi_i^\mu,k} ] = \frac{a}{S}$, it is straightforward
to compute the averages (see Eq. (\ref{hui}))
\begin{equation}
    \mathbb{E}_\xi[ \eta_{i\mu}^k ] = 0, \quad
 \mathbb{E}_\xi[ (\eta_{i\mu}^k)^2 ] = \frac{a}{S}\Big(1-\frac{a}{S}\Big) \quad \text{and}  \quad
 \mathbb{E}_\xi[ \eta_{i\mu}^k\eta_{i\mu}^l ] =  - \frac{a^2}{S^2}. \nonumber
\end{equation}
Substituting these results into Eq. (\ref{ew1}), we can rewrite $\mathcal{F}$ as
\begin{align}
 \mathcal{F} \simeq \;&\exp\Bigg({
    - i \int_0^T dt \frac1{N_a} 
    \left[ \sum_{i,k} \hat r_i^k(t) \eta_{ik}^1\right]
    \left[ \sum_{i,k} \sigma_i^k(t) \eta_{ik}^1\right]
    }\Bigg)\nonumber\\[3mm]  
    & \times \prod_{\mu=2}^p \prod_{i=1}^N
    \int Dx_{\mu} Dy_{\mu} D\hat{x}_{\mu} D\hat{y}_{\mu} 
    \;\;
    \Bigg[\exp\Bigg({i\int_0^Tdt \; 
    \Big[ 
    -x_\mu(t)y_\mu(t)
    + \hat x_\mu(t)x_\mu(t)
    + \hat y_\mu(t)y_\mu(t)
    \Big] }\Bigg) \nonumber \\
    &\times  
    \exp\Bigg(
    \frac{i}{N_a} \sum_{kl} C_{kl} U_i^{kl}
    - \frac{1}{2 N_a} \sum_{kl} C_{kl} V_{i\mu}^{k} V_{i\mu}^{l}
    \Bigg)\Bigg], \label{popo1}
\end{align}
where $\{ C_{kl} \}_{k,l=1}^S$ are the matrix elements that encode the color interactions
\begin{equation}
C_{kl} = \frac{a}{S}\Big(1-\frac{a}{S}\Big)\delta_{kl} 
- \frac{a^2}{S^2}(1-\delta_{kl}).
\end{equation}
Finally, we substitute the explicit forms of $U_i^{kl}$ and $V_{i\mu}^{k}$ back 
into Eq. (\ref{popo1}), integrate
over $x_{\mu}(t)$ and $y_{\mu}(t)$, and get the expression 
\begin{align}
&\mathcal{F} \simeq \exp\Bigg[
    - \frac{iN}{a(1-\frac{a}{S})} \int_0^T dt \sum_{kl}
    \Big(\frac{1}{N} \sum_{i} \eta_{ik}^1 \hat r_i^k(t) \Big)
    \Big(\frac{1}{N} \sum_{i} \eta_{il}^1 \sigma_i^l(t) \Big) 
    +\frac{i\alpha N}{a(1-\frac{a}{S})}\int_0^Tdt \sum_{kl} C_{kl} \Big(\frac{1}{N}\sum_{i}\hat r_i^k(t)\sigma_i^l(t)\Big)
    \Bigg] \nonumber \\
    &\times
    \exp\Bigg\{
    \alpha N \ln
    \Bigg[
    \int D\hat xD\hat y \;
    \exp\Big(i\int_0^Tdt \; 
    \hat x(t) \hat y(t)
    \Big) \times 
    \exp
    \Bigg(
    -\frac{1}{2a N(1-\frac{a}{S})}
    \int_0^T dt dt' \sum_{kl} C_{kl}
    \Big[
    \hat x(t) \hat x(t')  \sum_i \hat r_i^k(t) \hat r_i^l(t')   \nonumber \\
    & + 
    \hat y(t)  \hat y(t') \sum_i \sigma_i^k(t) \sigma_i^l(t')  + 2 \hat y(t) \hat x(t')
    \sum_i \sigma_i^k(t) \hat r_i^l(t') 
    \Big]
    \Bigg)
    \Bigg]
    \Bigg\}. \label{ytew}
\end{align}
The above equation provides the leading contribution arising from the disorder average of $\mathcal{Z}[\boldsymbol{\psi}]$ for large $N$.

The next step is to obtain a factorized form over sites by introducing the dynamical order-parameters
\begin{align}
    &m_k(t) = \frac{1}{N}\sum_{i}\eta_{ik}^1 \sigma_i^k(t) ,\\
    &l_k(t) = \frac{1}{N}\sum_{i}\eta_{ik}^1  \hat r_i^k(t) ,\\
    &Q_{kl}(t,t') = \frac{1}{N}\sum_i\hat r_i^k(t) \hat r_i^l(t') ,\\
    &K_{kl}(t,t')= \frac{1}{N}\sum_i\hat r_i^k(t)\sigma_i^l(t'),\\
    &q_{kl}(t,t') = \frac{1}{N}\sum_i\sigma_i^k(t)\sigma_i^l(t').
\end{align}
These order-parameters are introduced in Eq. (\ref{ytew}) through Dirac $\delta$-functionals, yielding
\begin{align}
\mathcal{F} &=   
    \int \Big( \prod_k D l_k D \hat l_k D m_k D \hat m_k \Big) 
    \int \Big( \prod_{kl} D Q_{kl} D \hat Q_{kl} D K_{kl} D \hat K_{kl} D q_{kl} D \hat q_{kl} \Big) \nonumber \\
    & \times \exp \Bigg(
    i \int_0^T dt \sum_k [\hat l_k(t) l_k(t) + \hat m_k(t) m_k(t)] 
    - \frac{iN}{a(1-\frac{a}{S})} \int_0^T dt \sum_{kl}
    l_{k}(t) m_{l}(t) 
    \Bigg) \nonumber \\
    &\times \exp \Bigg( i \int_0^T dt\;dt' \sum_{kl} [\hat Q_{kl}(t,t') Q_{kl}(t,t') + \hat K_{kl}(t,t') K_{kl}(t,t') + \hat q_{kl}(t,t') q_{kl}(t,t') ]
    \Bigg)   \nonumber \\
     &\times
    \exp\Bigg(
    \frac{i\alpha N}{a(1-\frac{a}{S})}\int_0^Tdt \sum_{kl} C_{kl} K_{kl}(t,t)
    + \alpha N \ln \mathcal{N} [\{Q_{kl},K_{kl},q_{kl}\}]
    \Bigg)  \nonumber \\    
    &\times \prod_i
    \exp\Bigg(
    -\frac{i}{N}\int_0^T dt \sum_k \eta_{ik}^1 [\hat l_k(t) \hat r_i^k(t) +\hat m_k(t) \sigma_i^k(t)] \Bigg) \nonumber \\
    & \times \exp \Bigg( - \frac{i}{N} \int dt\;dt' \sum_{kl} 
    [
    \hat Q_{kl}(t,t')\hat r_i^k(t)\hat r_i^l(t')
    +\hat K_{kl}(t,t')\hat r_i^k(t)\sigma_i^l(t')
    +\hat q_{kl}(t,t')\sigma_i^k(t)\sigma_i^l(t')
    ]
    \Bigg),    \label{uipo}
\end{align}
where the conjugate parameters $\{ \hat m_k, \hat l_k, \hat Q_{kl}, \hat K_{kl}, \hat q_{kl} \}$ arise
from the Fourier transforms of the associated $\delta$-functionals. The quantity $\mathcal{N} [\{Q_{kl},K_{kl},q_{kl}\}]$ is defined as
\begin{align}
    &\mathcal{N}[\{Q_{kl},K_{kl},q_{kl}\}] = 
    \int D\hat xD\hat y \;
    \;\;
    \exp\Bigg({i\int_0^Tdt \; 
    \hat x(t) \hat y(t)
    } \Bigg)\nonumber\\
    &\times
    \exp
    \Bigg(
    -\frac{1}{2a(1-\frac{a}{S})}
    \int_0^T dt dt' \sum_{kl} C_{kl}
    \Big[
    \hat x(t)Q_{kl}(t,t')\hat x(t')
    + 
    \hat y(t)q_{kl}(t,t')\hat y(t') 
    + 2 \hat y(t)K_{kl}(t,t')\hat x(t')
    \Big]
    \Bigg).
\end{align}
Inserting Eq. (\ref{uipo}) into the generating functional, Eq. (\ref{eq: GF}), and rescaling the conjugate 
order-parameters with $N$, we arrive at the compact form,
\begin{equation}
\overline{Z[\psi]} \simeq \int D \mathcal{G} \, e^{N \Phi[\mathcal{G}]},
\label{rem}
\end{equation}
with 
\begin{equation}
\mathcal{G} = \{ l_k,\hat l_k, m_k,\hat m_k, Q_{kl},
\hat Q_{kl}, K_{kl}, \hat K_{kl}, q_{kl},\hat q_{kl}  \}_{k,l=1}^{S}
\nonumber
\end{equation}
denoting the set of all functional order-parameters. The effective action per site
is given by

\begin{align}
    \Phi[&\mathcal{G}] =  
    i \int_0^T dt \sum_k [\hat l_k(t) l_k(t) + \hat m_k(t) m_k(t)] \nonumber \\
    &+ i \int_0^T dt\;dt' \sum_{kl} [\hat Q_{kl}(t,t') Q_{kl}(t,t') + \hat K_{kl}(t,t') K_{kl}(t,t') + \hat q_{kl}(t,t') q_{kl}(t,t') ] \nonumber \\
    &-  \frac{i}{a(1-\frac{a}{S})} \int_0^T dt \sum_{kl}
    l_k (t) m_l (t) 
    +\frac{i\alpha}{a(1-\frac{a}{S})}\int_0^Tdt \sum_{kl} C_{kl} K_{kl}(t,t)
    + \alpha \ln \mathcal{N} [\{Q_{kl},K_{kl},q_{kl}\}] \nonumber  \\
    &+ \frac{1}{N}\sum_i \ln \Bigg\{
    \int \Big(\prod_k Dr^k D\hat r^k \Big) 
    \exp\Bigg(
    i \int_0^T dt \sum_k \hat r^k(t) \left[ \tau_r \dot r^k(t) + \Omega^k(t) + r^k(t) \right] 
    \Bigg) \nonumber \\
    &\times
    \exp\Bigg(
    - i\int_0^T dt \sum_k \eta_{ik}^1 \, [\hat l_k(t)\hat r^k(t) +\hat m_k(t) \sigma^k(t)]
    + i\int_0^T dt \sum_k \psi_i^k(t) r^k(t) \Bigg) \nonumber  \\
    &\times \exp \Bigg( - i \int dt  dt' \sum_{kl} 
    [
    \hat Q_{kl}(t,t')\hat r^k(t)\hat r^l(t')
    +\hat K_{kl}(t,t')\hat r^k(t)\sigma^l(t')
    +\hat q_{kl}(t,t')\sigma^k(t)\sigma^l(t')
    ]  
    \Bigg)
    \Bigg\}. \label{sao}
\end{align}
Equations (\ref{rem}) and (\ref{sao}) constitute the main result of this section, as they recast the generating functional as an integral over the set $\mathcal{G}$ of macroscopic order parameters. In the next section, we evaluate this integral in the limit $N \to \infty$.

\subsection{The saddle-point equations}

In the limit \(N \rightarrow \infty\), the integrand in Eq. (\ref{rem}) is dominated by the stationary points of the action $\Phi[\mathcal{G}]$.
Setting $\psi_i(t)=0$ and taking the functional derivatives of $\Phi[\mathcal{G}]$ with respect to
$\{ \hat l_k,  \hat m_k, \hat Q_{kl}, \hat K_{kl}, \hat q_{kl}  \}_{j,l=1}^{S}$, we obtain the saddle point equations
\begin{align}
    & l_k(t) = \lim\limits_{N \rightarrow \infty} \frac{1}{N} \sum_i \eta_{ik}^1 \langle \hat r^{k}(t) \rangle_{\text{eff}}^{(i)}, \label{yu1} \\
    & m_k(t) = \lim\limits_{N \rightarrow \infty} \frac{1}{N} \sum_i \eta_{ik}^1 \langle \sigma^{k}(t) \rangle_{\text{eff}}^{(i)}. \\
 & Q_{kl}(t,t') =\lim\limits_{N \rightarrow \infty}  \frac{1}{N} \sum_i \langle \hat r^{k}(t) \hat r^{l}(t')\rangle_{\text{eff}}^{(i)}, \\
    & K_{kl}(t,t') =\lim\limits_{N \rightarrow \infty}  \frac{1}{N} \sum_i \langle \hat r^{k}(t) \sigma^{l}(t')\rangle_{\text{eff}}^{(i)}, \\
    & q_{kl}(t,t') = \lim\limits_{N \rightarrow \infty} \frac{1}{N} \sum_i \langle \sigma^{k}(t) \sigma^{l}(t')\rangle_{\text{eff}}^{(i)}, \label{yu5}  
\end{align}
where $\langle \dots \rangle_{\text{eff}}^{(i)}$ represents the following average
over the microscopic fields $\{ r^k, \hat{r}^k \}$ 
\begin{equation}
  \langle \dots \rangle_{\text{eff}}^{(i)} = \frac{
  \int \Big( \prod\limits_{k=1}^S D r^k  D \hat{r}^k \Big) (\dots) e^{i \int_{0}^T dt 
  \sum_{k=1}^S  \hat{r}^k(t) \big[ \tau_r \dot{r}^k(t) + \Omega^{k}(t)  + r^k(t)    \big]  } W_i[\{ r^k, \hat{r}^k \}]
  }{ \int 
  \Big(\prod\limits_{k=1}^S D r^k  D \hat{r}^k \Big) e^{i \int_{0}^T dt 
  \sum_{k=1}^S  \hat{r}^k(t) \big[ \tau_r \dot{r}^k(t) + \Omega^{k}(t)  + r^k(t)    \big]  } W_i[\{ r^k, \hat{r}^k \} ]  }, \label{aver}
\end{equation}
with the site-dependent weight
\begin{align}
 &W_{i}[\{ r^k, \hat{r}^k \}] = \exp{\left( - i \int_{0}^{T} dt \sum_{k}  \eta_{ik}^{1}
 \left[ \hat  l_k(t) \hat{r}^k(t) + \hat  m_k(t) \sigma^{k}(t) \right] \right)} \nonumber \\
 & \times \exp{\left( - i \int_{0}^{T} dt dt' \sum_{k l}
 \left[   \hat{r}^k(t)   \hat Q_{kl}(t,t')  \hat{r}^l(t')  +  \hat{r}^k(t)   
 \hat K_{kl}(t,t')  \sigma^l(t')  +  \sigma^k(t)   \hat q_{kl}(t,t')  \sigma^l(t')  \right]
 \right)}. \label{trd}
\end{align}
Two remarks are in order. First, although not explicit in our notation, the activity $\sigma^{k}(t)$ depends on the fields $\{ r^{l}(t) \}_{l=1}^{S}$ through Eq. (\ref{act}). Second, the  
weight $W_{i}$ depends on the site
index $i$ only through the condensed pattern components $\{ \eta_{ik}^{1} \}_{k=1}^{S}$. Equations 
(\ref{yu1}-\ref{yu5}), however, will depend only on a finite number of effective single site dynamics as explained in appendix \ref{app: 2cond_K}. 
Taking the derivatives of $\Phi[\mathcal{G}]$ with respect to
$\{ l_k,  m_k, Q_{kl}, K_{kl}, q_{kl} \}_{j,l=1}^{S}$, we obtain the saddle point equations for the conjugate order-parameters
\begin{align}    
    & \hat l_k(t) = \frac{1}{a(1-\frac{a}{S})} \sum_l m_{l}(t), \label{po1} \\
    & \hat m_k(t) = \frac{1}{a(1-\frac{a}{S})} \sum_l l_{l}(t), \label{po2} \\
    &\hat Q_{kl}(t,t') = \frac{-i\alpha \; C_{kl}}{2a(1-\frac{a}{S})} 
    \frac{\int D\hat x D\hat y \, \hat x(t) \, \hat x(t')  F [\hat x, \hat y  ]  }
    {\int D\hat x D \hat y \, F [\hat x, \hat y  ] }, \label{po3} \\
    &\hat K_{kl}(t,t') = \frac{-i\alpha \; C_{kl}}{a(1-\frac{a}{S})} 
    \frac{\int D\hat x D\hat y \; \hat x(t) \hat y(t') 
    F [\hat x, \hat y  ] }{\int D\hat x D \hat y F [\hat x, \hat y  ] }  - \frac{\alpha C_{kl}}{a(1-\frac{a}{S})}\delta(t-t'), \label{po4} \\
    &\hat q_{kl}(t,t') = \frac{-i\alpha \; C_{kl}}{2a(1-\frac{a}{S})} 
    \frac{\int D\hat x D\hat y \, \hat y(t) \, \hat y(t') F [\hat x, \hat y  ] }{\int D\hat x 
    D \hat y \, F [\hat x, \hat y  ], \label{po5}
}
\end{align}
with 
\begin{align}
 F [\hat x, \hat y  ] &=  \exp \left( i \int_{0}^{T} dt \hat{x}(t) \hat{y}(t) \right)  \nonumber \\
  &\times \exp \Bigg( - \frac{1}{2a(1-\frac{a}{S})}
    \int_0^T dt dt' \sum_{kl} C_{kl}
    \Big[
    \hat x(t)Q_{kl}(t,t')\hat x(t') +  \hat y(t)q_{kl}(t,t')\hat y(t')
    + 2 \hat x(t)K_{kl}(t,t')\hat y(t')
    \Big] \Bigg). \label{ghgh}
\end{align}
Equations (\ref{po1}-\ref{po5}) determine the weight $W_{i}[\{ r^k, \hat{r}^k \}]$, which governs
both the effective microscopic dynamics of the model and the evolution of the macroscopic 
order parameters defined by Eqs. (\ref{yu1}-\ref{yu5}). Next, we simplify the saddle-point
equations together with the expression for $W_{i}[\{ r^k, \hat{r}^k \}]$.

From the normalization $\mathcal{Z}[0]=1$ of the generating functional, one can
show that $\langle \hat r^k(t) \rangle_{\text{eff}}^{(i)} = 0$  and $\langle \hat r^k(t) \hat r^l(t')\rangle_{\text{eff}}^{(i)} = 0$, implying that 
\begin{equation}
    l_k(t)= \hat m_k(t) = 0 \quad \text{and} \quad Q_{kl}(t,t')=0.
\end{equation}
  The macroscopic parameter $m_k(t)$ is the overlap with the condensed pattern, $q_{kl}(t,t')$ is the auto-correlation between the activities, and $K_{kl}(t,t')$ is directly related to the response function $G_{kl}(t, t') = \frac{\delta\langle  \sigma^k(t)  \rangle_{\text{eff}}^{(i)}}{\delta \tilde\theta^l(t')}
  \Big{|}_{\tilde\theta=0}$ of the activity $\sigma^k(t)$ at time $t$ due to a perturbation $\tilde\theta^l(t')$ at time $t' < t$ as follows
  \begin{equation}
      K_{kl}(t,t') = i G_{lk}(t',t).
   \end{equation}
Hence, setting $Q_{kl}(t,t')=0$ in Eq. (\ref{ghgh}), we explicitly evaluate the Gaussian integrals 
in Eqs. (\ref{po3}-\ref{po5}) and obtain the conjugate order parameters
\begin{align}
    \hat Q_{kl}(t,t') & = \frac{-i\alpha \; C_{kl}}{2a(1-\frac{a}{S})} 
\Big[
(\mathbb{I}-\tilde G)^{-1} \tilde q (\mathbb{I}-\tilde G^{T})^{-1} 
\Big] (t, t'), \label{eqQ} \\
\hat q_{kl}(t,t') & = 0, \\
\hat K_{kl}(t,t') & =  \frac{\alpha \; C_{kl}}{a(1-\frac{a}{S})}
\Big[
\tilde G (\mathbb{I} - \tilde G)^{-1}
\Big](t,t'), \label{eqK}
\end{align}
where the colour-independent operators $\tilde q(t,t')$ and $\tilde G(t,t')$ are given by
\begin{align}
    \tilde q(t,t') = \frac{1}{a(1-\frac{a}{S})} \sum_{kl} C_{kl} q_{kl}(t,t'),  \\
    \tilde G(t,t') = \frac{1}{a(1-\frac{a}{S})} \sum_{kl} C_{kl} G_{kl}(t,t').
\end{align}
In the above equations, $\mathbb{I}(t,t')= \delta(t-t')$ is the identity operator, while
$\tilde G^{T}(t,t') = \tilde G(t',t)$ denotes the transpose of $\tilde G$.

\subsection{The effective single-site equation} 

The final step is to derive the effective single-site stochastic equation for the microscopic
field $r^{k}(t)$. This equation governs the evolution of the overlap $m_k(t)$, the correlation
of activities $q_{kl}(t,t')$, and the response function $G_{kl}(t,t')$ through the effective average
$\langle \dots \rangle_{\rm eff}^{(i)}$ defined in Eq. (\ref{aver}). Substituting the order parameters obtained in the previous section into Eq. (\ref{trd}), we obtain
a simplified form of the weight $W_{i}[\{ r^k, \hat{r}^k \}]$,
\begin{align}
 W_{i}[\{ r^k, \hat{r}^k \}] &= \exp{\left( - i \int_{0}^{T} dt \sum_{k} \hat{r}^k(t) \left[ \eta_{ik}^{1}
  \hat  l_k(t) + \int_{0}^{T} dt'  \sum_{l} \hat K_{kl}(t,t')  \sigma^l(t')    \right]  \right)} \nonumber \\
 & \times \exp{\left( - i \int_0^T\int_{0}^{T} dt dt' \sum_{k l} \hat{r}^k(t)   \hat Q_{kl}(t,t')  \hat{r}^l(t')   
 \right)}, \label{trdp}
\end{align}
where $\hat  l_k(t)$, $\hat Q_{kl}(t,t')$ and $\hat K_{kl}(t,t')$ are given, respectively, by Eqs. (\ref{po1}), (\ref{eqQ}) and (\ref{eqK}). The effective stochastic equation for $r^k(t)$ can
be inferred from the average activity $\langle \sigma^{k} (t) \rangle_{\rm eff}^{(i)}$. Since 
$\sigma^{k} (t)$ is independent of the conjugate 
fields $\{ \hat{r}^k(t) \}_{k=1}^{S}$, Eq. (\ref{aver}) yields
\begin{align}
   \langle \sigma^{k} (t) \rangle_{\rm eff}^{(i)} = 
   \int \Big( \prod_{l=1}^{S} D r^{l} \Big) \sigma^{k} (t) \mathcal{P}_{\rm eff}^{(i)}[\{ r^l \}],
\end{align}
where
\begin{equation}
\mathcal{P}_{\rm eff}^{(i)}[\{ r^k \}] = \frac{
  \int \Big( \prod\limits_{k=1}^S  D \hat{r}^k \Big)  e^{i \int_{0}^T dt 
  \sum_{k=1}^S  \hat{r}^k(t) \big[ \tau_r \dot{r}^k(t) + \Omega^{k}(t)  + r^k(t)    \big]  } W_i[\{ r^k, \hat{r}^k \}]
  }{ \int 
  \Big(\prod\limits_{k=1}^S D r^k  D \hat{r}^k \Big) e^{i \int_{0}^T dt 
  \sum_{k=1}^S  \hat{r}^k(t) \big[ \tau_r \dot{r}^k(t) + \Omega^{k}(t)  + r^k(t)    \big]  } W_i[\{ r^k, \hat{r}^k \} ]  }
\end{equation}
is the effective path-probability of 
$\{ r^k(t) \}_{k=1}^{S}$ in the limit $N \rightarrow \infty$.
Substituting  Eq. (\ref{trdp}) into the above equation and integrating 
over $\{ \hat r^k(t) \}_{k=1}^{S}$, we conclude that the effective stochastic dynamics
is driven by a temporally correlated noise. 
$W_i[\{r^k,\hat r^k\}]$ still depends on the site index $i$, through the
local configuration of the condensed patterns (in the general case with $B$ condensed pattern),
$(\xi^1_i,\dots,\xi^B_i)$, which enters via $\eta^1_{ik},\dots,\eta^B_{ik}$.
However there are only a finite number of distinct dynamical equations, and the reason is that if a permutation $\pi$ of the active
states (with $\pi(0)=0$) maps the condensed-pattern configuration of site $i$
onto that of site $j$, i.e.\ $\xi^\mu_j=\pi(\xi^\mu_i)$ for
$\mu=1,\dots,B$, then the color component $k$ of site $i$ and the component
$\pi(k)$ of site $j$ receive the same condensed-pattern fields and obey the
same effective equation. The derivation of the total number of distinct dynamical equations with $B$ condensed pattern and $S$ active colors is given in Appendix \ref{app: 2cond_K}

Units can therefore be grouped into $M(B,S)$ classes $g$ 
according to the realization of ($\xi_i^1,\dots,\xi_i^B$) up to a permutation of the active Potts colors.
The single-site field at colour $k$ of a unit in the class $g$ evolves in time according to the stochastic equation

\begin{equation}
   \tau_r \dot r_g^k(t) = - r_g^k(t) - \Omega_g^k(t) + \frac{\eta_{gk}^{1}}{a(1-\frac{a}{S})} \sum_{l=1}^{S} m_{l}(t) + \frac{1}{a(1-\frac{a}{S})}  \int_0^t dt'\sum_{l=1}^{S} C_{kl}
\mathcal{K}(t,t')\sigma_g^l(t') +\phi^k(t),
\end{equation}
with 
\begin{equation}
\mathcal{K}(t,t') \equiv \alpha \Big[\tilde G(\mathbb I-\tilde G)^{-1}\Big](t,t').
\end{equation}
The temporal correlated noise $\phi^{k}(t)$ is drawn from a Gaussian path-probability
with covariance 
\begin{equation}\label{eq: Gaussian_cov_cross_talk}
    \langle \phi^k(t) \phi^l(t') \rangle = \frac{\alpha C_{kl}}{a(1-\frac{a}{S})}
\Big[
(\mathbb{I}-\tilde G)^{-1} \tilde q (\mathbb{I}-\tilde G^{T})^{-1} 
\Big] (t, t').
\end{equation}
 
 The extension of the previous calculation to the case in which a finite number $B$ of condensed
 patterns is kept fixed, while the remaining $p-B$ patterns are 
 treated as quenched disorder, is straightforward. 
 The only modification appears in the drift term, which acquires an additional contribution 
 proportional to the overlaps with all condensed patterns. The resulting dynamics
 is governed by the stochastic equation
\begin{equation}\label{eq: single_site_eq_appendix}
    \tau_r \dot r_g^k(t) = - r_g^k(t) - \Omega_g^k(t) + \sum_{\mu=1}^{B} \frac{\eta_{gk}^\mu}{a(1-\frac{a}{S})} \sum_{l=1}^{S} m_{l}^{\mu}(t) + \frac{1}{a(1-\frac{a}{S})} \int_0^t dt'\sum_{l=1}^{S} C_{kl}
   \mathcal{K}(t,t')\sigma_g^l(t') +\phi^k(t).
\end{equation}
We remind the reader that $\Omega^k(t)$ is defined in Eq. (\ref{rewd}).

\section{Cavity method}\label{app: cavity_method} 

In this appendix, we derive the DMFT equations for the model using an alternative approach, namely the cavity method. We start from the explicit expression of the input received by state \(k\) of unit \(i\):
\begin{equation}
    h_i^k = \sum_{\substack{j=1 \\ j\neq i}}^N\sum_{m=1}^S J_{ij}^{km}\sigma_j^m.
\end{equation}
The connectivity matrix is fixed according the Hebb's rule
\begin{equation}\label{eq: J_matrix}
    J_{ij}^{km} = \frac{1}{Na(1-\frac{a}{S})}\sum_{\mu=1}^P\eta_{ik}^\mu\eta_{jm}^\mu,
\end{equation}
where $\eta_{ik}^\mu=\delta_{\xi_i^\mu,k}-\frac{a}{S}$.
Assuming there are two condensed patterns, we can separate the contribution of the condensed patterns in the input field as follows
\begin{equation}\label{eq: input}
        h_i^k(t) = \sum_{\mu=1}^2 \eta_{ik}^\mu m_\mu(t) +\frac{1}{Na(1-\frac{a}{S})}\sum_{\mu>2}\sum_{j(\neq i),m}\eta_{ik}^\mu\eta_{jm}^\mu\sigma_j^m(t),
\end{equation}
where $m_\mu(t)$ is the overlap of the activity of the network with the $\mu$ pattern:
\begin{equation}
    m_\mu(t) = \frac{1}{Na(1-\frac{a}{S})}\sum_{j,m}\eta_{jm}^\mu\sigma_j^m(t).
\end{equation}
In principle, the term that depends on unit $i$ should not be included in the right hand side of the above equation, but in the thermodynamic limit its effect is negligible.

The physical reason behind the fact that the average of the second term on the right hand side of Eq. (\ref{eq: input}), the cross-talk term, is different from zero in the $N \rightarrow \infty$ limit, is because there are correlations along loops in the system, and the number of stored patterns scales
with $N$. 
The difficulty in computing the average with respect to the pattern distribution comes from the fact that the activity of a unit $j$ of the network, $\sigma_j^m(t)$, is correlated with $J_{ij}^{km}$. The reason is that $\sigma_j^m(t)$ depends on the full history of unit $i$ through the connectivity $J_{ji}^{m'k}$, which is correlated with $J_{ij}^{km}$.

The disorder average of the cross-talk term comes from the contribution of all feedback loops of the network. In the large $N$ limit, we can consider the activity of unit $i$ in a time interval $[0,t]$ as a small perturbation to the activity of $\sigma_j^m(t)$. This perturbation propagates through all possible loops of the system before entering into $\sigma_j^m(t)$ (a simple scaling argument suggests that only loops where the unit $i$ appears once have a non-negligible contribution). Therefore, we can write the following expansion for $\sigma_j^m(t)$ 
\begin{equation}\label{eq: sigma_j_linear}
    \sigma_j^m(t) = \sigma_j^m(t)_{\backslash i} + \sum_{l=2}^{\infty}\delta^{(l)}\sigma_j^m(t),
\end{equation}
where the first term on the right hand side is the activity of unit $j$ when unit $i$ is removed from the network (this is the component of the activity which is independent of $J_{ij}^{km}$), and the second term is the linear response of unit $j$ caused by the past activity of unit $i$ that propagates through a loop of length $l$. 
For example, the linear response generated from  a loop of length 2, where the activity of unit $i$ is injected directly into unit $j$, is given by:
\begin{equation}
    \delta^{(2)}\sigma_j^m(t) = \int dt' \sum_{l,k=1}^S\frac{\delta \sigma_j^m(t)}{\delta \tilde\theta_j^l(t')}\; J_{ji}^{lk}\sigma_i^k(t') = \int dt' \sum_{l,n=1}^SG_j^{ml}(t,t')\; J_{ji}^{ln}\sigma_i^n(t').
\end{equation}
In the above equation, we have introduced the linear response function $G_j^{ml}(t,t')$, which tells how a unitary perturbation on $r_j^l(t')$ that is the preactivation of the state $l$ of site $j$ at time $t'$ will perturb the dynamics of $\sigma_j^m(t)$, the state $m$ on site $j$ at a later time $t$:
\begin{equation}
    G_j^{ml}(t,t') = \frac{\delta \sigma_j^m(t)}{\delta \tilde\theta_j^l(t')}.
\end{equation}

We can also write in full generality the expression for $\delta^{(l)}\sigma_j^m(t)$. The indices $i_2,i_3,\cdots i_l,i_{l+1}$ go from the last to the first node in the loop, so $i_2=j$, and $i_1=i_{l+1}=i$:
\begin{equation}
    \delta^{(l)}\sigma_j^m(t) = \int_0^tdt' \Bigg[\sum_{i_3,\cdots i_l=1}^N\;\; \sum_{m_2,\cdots m_l=1}^S\;\; \sum_{m'_1, \cdots m'_l=1}^S \; \prod_{\alpha=2}^l\; G_{i_\alpha}^{m'_{\alpha-1}m_{\alpha}}J_{i_\alpha i_{\alpha+1}}^{m_\alpha m_\alpha'}\;\delta_{m_1',m}\Bigg](t,t')\;\sigma_i^{m'_l}(t')
\end{equation}
Where the product of response function is a convolution:
\begin{equation}
    \Bigg[G_{i_2}^{m'_1m_2}G_{i_3}^{m'_2m_3}\Bigg](t,t') = \int_0^tdt''G_{i_2}^{m'_1m_2}(t,t'')G_{i_3}^{m'_2m_3}(t'',t').
\end{equation}

We can now insert \ref{eq: sigma_j_linear} inside the cross-talk term in equation  \ref{eq: input}, and compute the disordered average. 

The underlying assumption in the calculation below is that we are considering the input that unit $i$ would feel on average at time $t$, caused by its influence on the activity of the whole network, neglecting eventually the correlation of unit $i$ and the weights along the loop. This is equivalent to neglect the contribution of loops that come back onto unit $i$ a number of time $n\geq1$. We can neglect this contribution because the contribution from all the connection weights along a loop of length $l$ is of $O(N^{-l})$, while the number of loops of length $l$ where $i$ appears $n$ times is of $O(N^{l-n-1})$. The average from all the sums over the pattern will contribute only with a term of $O(\alpha N)$ because they are chosen to be independent, so these loops contribute to the input with a term of $O(N^{-n})$, which is negligible for $n\geq 1$ as $N \rightarrow \infty$.
Hence, only the contributions from loops where unit $i$ appears at the beginning and at the end of the loop are of $O(1)$, with a prefactor proportional to $\alpha$.
Both the correlations along loops and the extensive number of patterns is thus required for having a non zero average of the cross-talk term.


We can now use the explicit expression of the connectivity matrix as in \ref{eq: J_matrix} inside \ref{eq: sigma_j_linear} and plug everything inside \ref{eq: input}:
\begin{align}
h_i^k(t)
= \sum_{\mu=1}^2 \eta_{ik}^\mu m_\mu(t)
&+ \sum_{l=2}^{\infty}\frac{1}{N_a^l}
\int_0^t dt'\,\Bigg\{
\sum_{\mu_1,\mu_l=1}^{P}\sum_{m_l'=1}^S
\left\langle
\eta_{ik}^{\mu_1}\eta_{im'_l}^{\mu_l}
\right\rangle
\nonumber\\[0.2cm]
&\quad\times
\Bigg[
\prod_{\alpha=2}^l
\sum_{i_\alpha=1}^N
\sum_{\mu_{\alpha-1},\mu_\alpha=1}^{ P}\sum_{m'_{\alpha-1},m_\alpha=1}^S
\left\langle
\eta_{i_\alpha m'_{\alpha-1}}^{\mu_{\alpha-1}}
\eta_{i_\alpha m_\alpha}^{\mu_\alpha}
\right\rangle
G_{i_\alpha}^{m'_{\alpha-1}m_\alpha}
\Bigg](t,t')\,
\sigma_i^{m'_l}(t')\Bigg\}.
\end{align}
where $N_a=Na(1-a/S)$

The average over the patterns is computed from
\begin{equation}
    \langle\eta_{ik}^\mu\eta_{jl}^{\mu'} \rangle =\delta_{ij}\delta_{\mu\mu'}C_{kl} =\delta_{ij}\delta_{\mu\mu'}\Bigg(\frac{a}{S}\delta_{kl}-\frac{a^2}{S^2}\Bigg).
\end{equation}
In addition we can also define the average response function,
\begin{equation}
    G_{kl}(t,t')=\frac{1}{N}\sum_{i=1}^NG_i^{kl}(t,t'),
\end{equation}
and the projected response function,
\begin{equation}
    \tilde G(t,t')=\frac{1}{a(1-\frac{a}{S})}\sum_{k,l=1}^SC_{kl}G_{kl}(t,t').
\end{equation}
Because the patterns are statistically independent, the sums over the patterns contribute with a factor $\alpha N$, and the final result for the disorder average is given by 
\begin{align}
     h_i^k(t) &= \sum_{\mu=1}^2 \eta_{ik}^\mu m_\mu(t)+ \int_0^tdt'\sum_{l=2}^{\infty} \Bigg[\frac{\alpha}{a(1-\frac{a}{S})}\sum_{m=1}^SC_{km}(\tilde G)^{l-1}
     \Bigg](t,t')\sigma_i^{m}(t')\nonumber\\
     &=\sum_{\mu=1}^2 \eta_{ik}^\mu m_\mu(t)+\int_0^tdt'\frac{1}{a(1-\frac{a}{S})}\sum_{m=1}^SC_{km}\mathcal{K}(t,t')\sigma_i^m(t'),
\end{align}
where we have defined the memory kernel
\begin{equation}
    \mathcal{K}(t,t') = \alpha\Big[\tilde G(\mathbb I-\tilde G)^{-1}\Big](t,t').
\end{equation}
The same expansion in Eq. (\ref{eq: sigma_j_linear}) can be used to compute the covariance of the cross-talk term, between the input coming inside $\sigma_i^k(t)$ and $\sigma_i^l(t')$, and it will give the same result of Eq. (\ref{eq: Gaussian_cov_cross_talk}).

\section{Number of pattern configurations and distinct dynamical equations}\label{app: 2cond_K} 

We define a site class as an equivalence class of local condensed-pattern configurations under permutations of the active Potts states. Thus, two sites \(i\) and \(j\) belong to the same class if there exists a permutation \(\pi\) of the active states, with \(\pi(0)=0\), such that
$\xi_j^\mu=\pi(\xi_i^\mu)$ ($\mu=1,\ldots,B$).
The permutation \(\pi\) is a bijection on the active states, and therefore maps distinct active colors to distinct active colors.

When this condition holds, the same permutation maps the color components of site \(i\) one-to-one onto those of site \(j\). Corresponding components then receive the same condensed-pattern fields (that is the term proportional to $\eta_k^\mu$ in Eq. \ref{eq: single_site_eq_appendix}) and are coupled, through the memory kernel \(\mathcal K\), to the same set of color components. Therefore, the two sets of color components obey identical effective dynamical equations, provided that their initial conditions are also mapped consistently. If no such permutation exists, the corresponding dynamics is different, and the two sites cannot be included in the same class.

We can now count the number of distinct single-site equations that have to be evolved. We first count the number of possible realizations of $(\xi_i^1,\ldots,\xi_i^B)$
up to a permutation $\pi$ of the active Potts colors, with $\pi(0)=0$. This number is
\begin{equation}\label{eq:distinct_g}
M(B,S)
=
\sum_{r=0}^{B}
\binom{B}{r}
\sum_{m=0}^{\min(r,S)}
\left\{ {r \atop m} \right\}.
\end{equation}
Here \(r\) is the number of condensed patterns which are active at the site, while \(m\) is the number of distinct active Potts colors appearing among these \(r\) patterns. The binomial factor \(\binom{B}{r}\) selects which of the \(B\) condensed patterns are active. Once these \(r\) patterns have been selected, the Stirling number $\left\{ {r \atop m} \right\}$ of the second kind counts the number of ways of partitioning them into \(m\) non-empty groups of equal color. The \(r\) active states are labeled (by \(\mu\)) objects, since each of them is associated with its own macroscopic overlap \(m^\mu(t)\), and in general \(m^\mu(t)\neq m^\nu(t)\). By contrast, the \(m\) groups are indistinguishable, because they represent active colors only up to a permutation of the Potts states. For instance, for \(B=3\), \(r=3\) and \(m=2\), the configurations \(xxy\) and \(yyx\) represent the same class after the permutation \(x\leftrightarrow y\).

The number \(M(B,S)\) counts site classes, but not yet the number of distinct equations that we have. Indeed, within a given site, one must still specify which color component is being evolved. If the condensed pattern configuration contains \(m\) distinct active colors, the components for each of these \(m\) colors will follow different dynamics. In addition, if \(m<S\), all active colors different from the condensed ones are equivalent by symmetry and can be represented by one additional component, denoted by \(k\). Thus, a site with \(m\) distinct condensed colors gives rise to $m+\mathbf{1}_{m<S}$ distinct components.

Therefore, the total number of distinct effective equations is obtained by weighting each local condensed-pattern configuration by this number of representative components:
\begin{equation}
N_{\mathrm{cl}}(B,S)
=
\sum_{r=0}^{B}
\binom{B}{r}
\sum_{m=0}^{\min(r,S)}
\left\{ {r \atop m} \right\}
\left(
m+\mathbf{1}_{m<S}
\right).
\end{equation}
For two fixed condensed memories, each site is characterized by the local pair $(\xi_i^1,\xi_i^2)$. Identifying configurations that differ only by a relabeling of the active Potts colors, one obtains five different local configuration classes: $xx,\; xy,\; xs,\; sy,\; ss$, for $S\ge2$.

Within each of these site classes, one must still specify which color component \(k\) is being considered. The component can either correspond to one of the active colors appearing in the condensed-pattern configuration, or to an active color different from all the condensed colors present at that site. Denoting this generic non-condensed active color by \(k\), the five site classes generate the following $10$ components: \(xxx,\, xxk,\, xyx,\, xyy,\, xyk,\, xsx,\, xsk,\, syy,\, syk,\, ssk\), if $S\ge3$.

\section{Derivation of $xyy$, $syy$ equations}\label{app: xyy syy equations} 

For $\alpha=0$, we can write an explicit form for the magnetization term in the drift of the mean field equation \ref{eq: single site eq}, keeping in mind that all the components of a unit in a color different from the two condensed ones (e.g. $\sigma_{xy}^k(t)$) will evolve with the same dynamics. (In the following we use $x$ and $y$ for the two condensed colors and $k$ for colors different 
from $x$ and $y$).
The overlap with the first and second patterns is given by
\begin{align}
m^1(t) &= \frac{1}{N_a}\Bigg\{
\begin{aligned}[t]
&C_1\left[\left(1-\frac{a}{S}\right)\sigma_{xx}^x
-\frac{a(S-1)}{S}\sigma_{xx}^k\right]
+C_2\left[\left(1-\frac{a}{S}\right)\sigma_{xy}^x
-\frac{a}{S}\sigma_{xy}^y
-\frac{a(S-2)}{S}\sigma_{xy}^k\right]\\
&+(Na-C_1-C_2)\left[
\left(1-\frac{a}{S}\right)\sigma_{xs}^x
-\frac{a(S-1)}{S}\sigma_{xs}^k
-\frac{a}{S}\sigma_{sy}^y
-\frac{a(S-1)}{S}\sigma_{sy}^k\right]\\
&-\bigl[N(1-2a)+C_1+C_2\bigr]a\sigma_{ss}^k
\Bigg\}
\end{aligned}
\label{eq: m1}\\[0.5em]
m^2(t) &= \frac{1}{N_a}\Bigg\{
\begin{aligned}[t]
&C_1\left[\left(1-\frac{a}{S}\right)\sigma_{xx}^x
-\frac{a(S-1)}{S}\sigma_{xx}^k\right]
+C_2\left[\left(1-\frac{a}{S}\right)\sigma_{xy}^y
-\frac{a}{S}\sigma_{xy}^x
-\frac{a(S-2)}{S}\sigma_{xy}^k\right]\\
&+(Na-C_1-C_2)\left[
-\frac{a}{S}\sigma_{xs}^x
-\frac{a(S-1)}{S}\sigma_{xs}^k
+\left(1-\frac{a}{S}\right)\sigma_{sy}^y
-\frac{a(S-1)}{S}\sigma_{sy}^k\right]\\
&-\bigl[N(1-2a)+C_1+C_2\bigr]a\sigma_{ss}^k
\Bigg\}
\end{aligned}
\label{eq: m2}
\end{align}
To better understand which mechanism drives the latching transition we will focus on the dynamics of the units $xy$ and $sy$ in the color $y$ of the second condensed pattern. When the network is aligned with the first pattern, these units are respectively active and silent with the same color of the first patter. During a latching event the $y$ component of these units will start to rise. We want to derive directly from their dynamical equations the boundary in the $C_1,C_2$ plane where latching is possible. 
\begin{eqnarray}
    { \dot{r}_{xy}^y}&=&h_{xy}^y-\theta_{xy}^y -r_{xy}^y +{w\over S}
    \left[ (S-1)\sigma_{xy}^y-\sigma_{xy}^x-(S-2)\sigma_{xy}^k\right]\nonumber\\
    {\dot{r}_{sy}^y}&=&h_{sy}^y-\theta_{sy}^y -r_{sy}^y +{w\over S}
    \left[ (S-1)\sigma_{syy}-(S-1)\sigma_{syk}\right] \label{twokeyfields}
\end{eqnarray}
with
\begin{eqnarray}\label{eq: keyfieldh}
    h_{xy}^y=h_{sy}^y&=&-{a\over S}m^1+\left(1-{a\over S}\right)m^2. 
\end{eqnarray}
Using Eq. \ref{eq: m1} and \ref{eq: m2} inside Eq. \ref{eq: keyfieldh} we obtain:
\begin{equation}
\begin{aligned}
-\frac{a}{S}m^1+\left(1-\frac{a}{S}\right)m^2
={}&k_{oy}^{\,y}\left[(1-X)Y\sigma_{xy}^{\,y}
+(1-Y)\sigma_{sy}^{\,y}\right]
+k_{xx}^{\,x}XY\sigma_{xx}^{\,x}\\
&+k_{xo}^{\,x}\left[(1-X)Y\sigma_{xy}^{\,x}
+(1-Y)\sigma_{xs}^{\,x}\right]
+k_{xo}^{\,k}\left[XY\sigma_{xx}^{\,k}
+(1-Y)\sigma_{xs}^{\,k}\right]\\
&+k_{sy}^{\,k}(1-Y)\sigma_{sy}^{\,k}
+k_{xy}^{\,k}(1-X)Y\sigma_{xy}^{\,k}
+k_{ss}^{\,k}\sigma_{ss}^{\,k}.
\end{aligned}
\end{equation}
where $X=C_1/(C_1+C_2)$ and $Y=(C_1+C_2)/Na$, and
\begin{eqnarray}
    k_{oy}^{\,y}&=&{a^2/S^2+(1-a/S)^2 \over (1-a/S)}\\
    k_{xx}^{\,x}&=&(1-2a/S)\\
    k_{xo}^{\,x}&=&-2a/S\\
    k_{xo}^{\,k}&=&-{a(S-1)(1-2a/S) \over S(1-a/S)}\\
    k_{sy}^{\,k}&=&-{a(S-1)(1-2a/S)\over S(1-a/S)}\\
    k_{xy}^{\,k}&=&-{a(S-2)(1-2a/S) \over S(1-a/S)}\\
    k_{ss}^{\,k}&=&-{a(1/a-2+Y)(1-2a/S) \over (1-a/S)}.
\end{eqnarray}
So finally the equations for the units $xyy$ and $syy$ are
\begin{eqnarray}
    { \dot{r}_{xy}^{\,y}}&=&M_{12}(\{\sigma'\})
    +k_{oy}^{\,y}\left[(1-X)Y{ \sigma_{xy}^{\,y}}+(1-Y){ \sigma_{sy}^{\,y}}\right]-\theta_{xy}^{\,y} -r_{xy}^{\,y}+{w\over S}
    \left[ (S-1){ \sigma_{xy}^{\,y}}-\sigma_{xy}^{\,x}-(S-2)\sigma_{xy}^{\,k}\right] \\
    {\dot{r}_{sy}^{\,y}}&=&M_{12}(\{\sigma'\})
    +k_{oy}^{\,y}\left[(1-X)Y{\sigma_{xy}^{\,y}}+(1-Y){\sigma_{sy}^{\,y}}\right]-\theta_{sy}^{\,y} -r_{sy}^{\,y}+{w\over S}
    \left[ (S-1){\sigma_{sy}^{\,y}}-(S-1)\sigma_{sy}^{\,k}\right], \label{twokeyeqs}
\end{eqnarray}
where $M_{12}$ is given by:
\begin{eqnarray}
M_{12}(\{\sigma'\})
&=&
k_{xx}^{\,x}XY\sigma_{xx}^{\,x}
+k_{xo}^{\,x}\left[(1-X)Y\sigma_{xy}^{\,x}+(1-Y)\sigma_{xs}^{\,x}\right]
\nonumber\\
&&
+k_{xo}^{\,k}\left[XY\sigma_{xx}^{\,k}+(1-Y)\sigma_{xs}^{\,k}\right]
+k_{sy}^{\,k}(1-Y)\sigma_{sy}^{\,k}
+k_{xy}^{\,k}(1-X)Y\sigma_{xy}^{\,k}
+k_{ss}^{\,k}\sigma_{ss}^{\,k}.
\label{eq:M12_def}
\end{eqnarray}
The notation $\{\sigma'\}$ indicates that $M_{12}$ depends on the
activities of all the Potts components except $\sigma^y_{xy}$ and
$\sigma^y_{sy}$.
If we now neglect the components with color $k$, we end up with the following expression:
\begin{equation}\label{app:eq_M_12}
M_{12}(\{\sigma'\}) \approx \left(1-\frac{2a}{S}\right) XY\sigma_{xx}^{\,x} -\frac{2a}{S} \left[(1-X)Y\sigma_{xy}^{\,x} + (1-Y)\sigma_{xs}^{\,x} \right].  
\end{equation}

\section{Numerical solution of the DMFT equations} \label{Numerics} 

We solve Eq. (\ref{eq: single site eq})  for
$B=2$ through a self-consistent iterative procedure. 
At each iteration, the two-time correlation $\tilde q(t,t')$ and response
$\tilde G(t,t')$ determine the memory kernel $K(t,t')$ and the covariance $D_{kl}(t,t')$ of the effective Gaussian noise according to Eqs. (\ref{eq: memory kernel}) and (\ref{eq: Gaussian noise}). 
These kernels are used to sample stochastic paths 
from Eq. (\ref{eq: single site eq}) independently for each 
pattern configuration $g$ or, equivalently, for each site class. The macroscopic 
order-parameters are subsequently 
updated by averaging over these stochastic paths with the corresponding weights of the pattern configurations.
The convergence of this iterative procedure is stabilized through 
adaptive damping, whose strength is dynamically adjusted according to the residual variation of the updated macroscopic observables across successive iterations. Below we outline the algorithm. 
\begin{center}
\fbox{\parbox{0.95\linewidth}{

\textbf{Algorithm for solving the DMFT equations}

\vspace{2mm}

\textbf{Input:}
initial guess for ($m^{1},m^{2},\tilde q,\tilde G$), damping 
parameter $\lambda\in(0,1]$, maximum number of iterations
$n_{\max}$, and convergence tolerance $\varepsilon$.

\vspace{2mm}

\textbf{For }
\(
n=1,\dots,n_{\max}
\):

\vspace{1mm}

\hspace*{3mm}
\textbf{(1) Construct the effective kernels}

\hspace*{6mm}
\(
K \leftarrow K[\tilde G]
\),

\hspace*{6mm}
\(
D \leftarrow D[\tilde{q}, \tilde{G}]
\).

\vspace{1mm}

\hspace*{3mm}
\textbf{(2) For each site class}
\(
g\in\{xx,xy,xs,sy,ss\}
\)

\hspace*{6mm}
Sample the Gaussian trajectories
\(
\phi_{g}^{k}(t)
\sim
\mathcal{N}(0,D)
\).

\hspace*{6mm}
Generate copies of the stochastic dynamics from Eq. (\ref{eq: single site eq}).

\hspace*{6mm}
Compute the conditioned order-parameters 
$(m_g^{1},m_g^{2},\tilde q_g,\tilde G_g)$ as averages over the stochastic trajectories.

\vspace{1mm}

\hspace*{3mm}
\textbf{(3) Recombine the observables}

\hspace*{6mm}
\(
X_{\mathrm{new}}
=
\sum_g c_g X_g,
\qquad
X\in
\{
m^{1},
m^{2},
\tilde q,
\tilde G
\}.
\)

\vspace{1mm}

\hspace*{3mm}
\textbf{(4) Update the order-parameters with damping $\lambda$}

\hspace*{6mm}
\(
X
\leftarrow
(1-\lambda)X
+
\lambda X_{\mathrm{new}}.
\)

\vspace{1mm}

\hspace*{3mm}
\textbf{(5) Convergence check}

\hspace*{6mm}
If $\max_X \|X_{\mathrm{new}}-X\|_\infty < \varepsilon$, then terminate.

\vspace{2mm}

\textbf{Output:}
macroscopic observables ($m^{1},m^{2},\tilde q,\tilde G$).

}}
\end{center}

\section{$\alpha =0$ phase diagram and the stability of the limit cycle}\label{app: limit cycle}
This appendix explains how a limit cycle of the dynamics is computed and how its stability is assessed. The periodic orbit is found using the shooting method \cite{press2007numerical}, and its stability is determined from the spectrum of the full monodromy matrix \cite{chicone2006ordinary} over one period.

\subsection{The dynamical system}
\label{app:system}
 
At $\alpha=0$ and $B=2$ the network reduces to the five site classes
$g\in\{xx,xy,xs,sy,ss\}$, each with $S$ Potts states, so the state of the system is the vector
\begin{equation}
v=\big(r^k_g,\;\theta^k_g\big)_{g=1,\dots,5;\;k=1,\dots,S}\in\mathbb R^{n},
\qquad n=2\times 5S.
\label{app:state}
\end{equation}
These are the $n$ unknowns: $n/2$ fields $r^k_g$ and $n/2$ adaptive thresholds $\theta^k_g$. 
The field of the silent state is fixed at $r^0=U$. The activities of the active states are
\begin{equation}
\sigma^k_g(r)=\frac{e^{\beta r^k_g}}{e^{\beta U}+\sum_{l=1}^{S}e^{\beta r^l_g}},
\qquad
\bar\sigma_g=\frac1S\sum_{l=1}^S\sigma^l_g ,
\label{app:softmax}
\end{equation}
and the overlaps and fields are
\begin{equation}
m^\mu(\sigma)=\frac{1}{a(1-a/S)}\sum_{g}c_g\sum_k \eta^\mu_{gk}\,\sigma^k_g,
\qquad
H^k_g(\sigma)=\sum_{\mu=1,2}\eta^\mu_{gk}\,m^\mu(\sigma)+w\big(\sigma^k_g-\bar\sigma_g\big),
\label{app:fields}
\end{equation}
with $c_{xx}=aXY$, $c_{xy}=a(1-X)Y$, $c_{xs}=c_{sy}=a(1-Y)$, $c_{ss}=1-2a+aY$ and
$\eta^\mu_{gk}=\delta_{k,\xi^\mu_g}-a/S$. The dynamics is the autonomous vector field
\begin{equation}
\dot v=F(v),\qquad
F(v)=\begin{pmatrix}\dfrac{-r+H(\sigma(r))-\theta}{\tau_r}\\[3mm]
\dfrac{\sigma(r)-\theta}{\tau_\theta}\end{pmatrix}.
\label{app:vectorfield}
\end{equation}
Its Jacobian is
\begin{equation}
DF(v)=\begin{pmatrix}\tau_r^{-1}\big(-\mathbb I+\mathcal H A\big) & -\tau_r^{-1}\mathbb I\\[1mm]
\tau_\theta^{-1}A & -\tau_\theta^{-1}\mathbb I\end{pmatrix},
\qquad
A_{gk,g'l}=\delta_{gg'}\beta\sigma^k_g\big(\delta_{kl}-\sigma^l_g\big),
\label{app:DF}
\end{equation}
\begin{equation}
\mathcal H_{gk,g'l}=\frac{\partial H^k_g}{\partial\sigma^l_{g'}}
=\frac{c_{g'}}{a(1-a/S)}\sum_{\mu=1,2}\eta^\mu_{gk}\eta^\mu_{g'l}
+w\,\delta_{gg'}\Big(\delta_{kl}-\frac1S\Big).
\end{equation}

We write $\Phi_t(v)$ for the solution of the dynamics at time $t$ with initial condition $\Phi_0(v)=v$; this is
the flow map of the system.

\subsubsection{Stability of the silent fixed point}
The silent fixed point $v^*$ is obtained by setting $\theta^k_g=\sigma^k_g$,
which reduces the fixed-point condition $F(v^*)=0$ to the $n/2$ equations
$r^k_g=H^k_g(\sigma(r))-\sigma^k_g(r)$. These are solved with Powell's hybrid
method (tolerance $10^{-12}$), starting from the low-activity guess
$r^k_g=U-0.5$. The solution is continued in $Y$, using the fixed point at the
previous value of $Y$ as initial guess.
It may occasionally converge to a different, active fixed point; such solutions are recognised by a non-vanishing overlap ($|m^1|>0.05$), and the solver is then restarted from the low-activity guess.
The linear stability of
$v^*$ is given by the eigenvalues $\lambda_i$ of the Jacobian
$DF(v^*)$, Eq.~\eqref{app:DF}, 
The silent state is stable when
$\max_i\mathrm{Re}\,\lambda_i<0$; the line $Y_H(X)$ in Fig.~\ref{fig:zero_cross_M12}
is the zero contour of $\max_i\mathrm{Re}\,\lambda_i$ on the $(X,Y)$ grid.
\subsection{Identification of different phases from finite size simulations}\label{app:numerical_details_phase}
To identify the different phases of figures \ref{fig:zero_cross_M12} and \ref{fig:alpha_diagram} from the finite size simulations we used the following criteria: each cell of the diagram is obtained by integrating the dynamics for
$40\,\tau_{\theta}$ from a cue on the first memory. We say that memory $\mu$ is retrieved when its overlap exceeds the
other by more than $0.05$ and the retrieval activity $A(t)=\max(m_1(t),m_2(t))$ is
above $0.3$, and we count a transition every time the retrieved memory changes.
With no transition the cell is labelled no latching. With a single transition it
is low latching if $A$ falls below $0.3$ between the two retrievals, that is if
the overlap with the first memory dies before the second is retrieved, and high latching
otherwise. With two or more transitions it is pathological: sustained if $A$,
averaged over the last $5\,\tau_{\theta}$ of the run, is still above $0.3$, and
transient otherwise. The resulting diagram is not sensitive to these precise tresholds: taking the
activity threshold anywhere between $0.2$ and $0.45$, and the minimum lead between $0.02$ and $0.1$, changes the phase of at most $4.5\%$ of the
cells.
 
\subsection{Shooting method to find periodic orbit}
\label{app:shooting}
 
A solution $v(t)$ of the ODE \eqref{app:vectorfield} is $T$-periodic, with $T>0$, if $v(t+T)=v(t)$ for all $t$. Given a point $v_0$ on the periodic curve, the orbit is the set $\Gamma=\{\Phi_t(v_0):0\le t<T\}$, a closed curve
in $\mathbb R^{n}$. Any given point $v$ on the orbit $\Gamma$ satisfies the following equation:
\begin{equation}
\Phi_T(v)-v=0.
\label{app:shoot_full}
\end{equation}
This gives $n$ scalar equations for the $n+1$ unknowns
$(v,T)\in\mathbb R^{n}\times\mathbb R$.
An additional equation is needed to fix the starting point on the cycle, and we choose it to be the point where the overlaps with two memories cross, so $v$ is required to lie on the hypersurface
\begin{equation}
\Sigma=\{v:\ D(v)\equiv m^1(v)-m^2(v)=0\}.
\label{app:section}
\end{equation}
The full system is
\begin{equation}
G(v_0,T)=\begin{pmatrix}\Phi_T(v_0)-v_0\\[1mm] m^1(v_0)-m^2(v_0)\end{pmatrix}=0
\in\mathbb R^{n+1},\qquad (v_0,T)\in\mathbb R^{n+1},
\label{app:shootsystem_full}
\end{equation}
which is solved by Newton's method: one starts from an initial guess for $(v_0,T)$, then solve $G(v_0,T)=0$, Eq.~\eqref{app:shootsystem_full}, by
\begin{equation}\label{eq:app_Newton}
\begin{pmatrix}v_0\\ T\end{pmatrix}^{(k+1)}=
\begin{pmatrix}v_0\\ T\end{pmatrix}^{(k)}-\Big[DG^{(k)}\Big]^{-1}G^{(k)}.
\end{equation}

\subsection{Stability: the monodromy matrix}
\label{app:floquet}
Let $\bar v(t)$ be the periodic solution and write
$v(t)=\bar v(t)+\delta v(t)$. To first order in $\delta v$,
\begin{equation}
\delta \dot v=DF\big(\bar v(t)\big)\,\delta v ,
\label{app:variational}
\end{equation}
this is a linear system whose coefficient matrix
$\mathcal J(t)=DF(\bar v(t))$ is $T$-periodic. Let $\Xi(t)$ be its principal fundamental matrix,
\begin{equation}
\dot\Xi(t)=\mathcal J(t)\,\Xi(t),\qquad \Xi(0)=\mathbb I ,
\label{app:fundamental}
\end{equation}
so that $\delta v(t)=\Xi(t)\,\delta v(0)$ for every initial perturbation. Because
$\Phi_t(\bar v(0)+\delta v)=\bar v(t)+\Xi(t)\delta v+O(\|\delta v\|^2)$, the fundamental matrix is
exactly the derivative of the flow with respect to the initial condition,
\begin{equation}
\Xi(t)=D\Phi_t(v_0).
\end{equation}
The monodromy matrix of the orbit is defined as the fundamental matrix evaluated after one full period:
\begin{equation}
\mathcal M_T=\Xi(T)=D\Phi_T(v_0).
\label{app:monodromy}
\end{equation}
Its eigenvalues $\rho_1,\dots,\rho_{n}$ are the Floquet multipliers. 
Since a perturbation tangent to the cycle is mapped onto itself after one period,
\begin{equation}
\mathcal M_T\,f_0=f_0,\qquad f_0=F(v_0),
\end{equation}
$\rho=1$ is always a Floquet multiplier of an autonomous system, with eigenvector the
tangent to the orbit. It expresses the freedom of sliding the
initial condition along the closed curve, which neither grows nor decays.
 
The orbit $\Gamma$ is asymptotically stable if the remaining $n-1$ multipliers all lie strictly inside the unit circle,
\begin{equation}
|\rho_i|<1\quad\text{for all }i\ \text{with}\ \rho_i\neq1 ,
\end{equation}
and unstable if at least one multiplier lies outside it.
It is possible to numerically evaluate $\mathcal M_T$ by differentiating the flow by finite differences:
 \begin{equation}
[\mathcal M_T]_{:,j}=\frac{\Phi_T(v_0+h e_j)-\Phi_T(v_0-h e_j)}{2h}.
\label{app:fdmono}
\end{equation} 
The critical line $Y_c$ in Fig.~\ref{fig:zero_cross_M12} is the locus where the
largest non-trivial Floquet multiplier of the limit cycle crosses the unit
circle. It is computed as follows.
\begin{enumerate}
\item \emph{Initial orbit.} At $(X,Y)=(0.90,0.98)$ the limit cycle is
attracting, so we integrate the dynamics from the cue until it reaches the
cycle, and take as initial guess the first point $v_0$ at which $m^1=m^2$,
with $T\simeq2\tau_\theta$. Newton's method,
Eq.~\eqref{app:shootsystem_full}, then gives the exact orbit.
\item \emph{Top row of the grid.} We move along the row $Y=0.98$ one grid point
at a time in $X$, starting from $X=0.90$. At each new $X$, Newton's method is
initialised with the orbit found at the previous grid point.
\item \emph{Each column.} For each $X$, starting from the orbit at $Y=0.98$, we
lower $Y$ step by step, again initialising Newton's method with the orbit found
at the previous value of $Y$.
\item \emph{Excluding the silent state.} The silent fixed point satisfies
Eq.~\eqref{app:shootsystem_full} for any $T$. We therefore accept a solution
only if its amplitude $\max_t|m^1-m^2|$ exceeds $0.02$.
\item \emph{Stability.} At each accepted solution $(v_0,T)$, i.e.\ at a point
$v_0$ on the periodic orbit, we compute the monodromy matrix
$\mathcal M_T=D\Phi_T(v_0)$ by finite differences, Eq.~\eqref{app:fdmono}, and
its eigenvalues. After removing the trivial multiplier $\rho=1$, the cycle is
stable if all remaining multipliers satisfy $|\rho_i|<1$.
\item \emph{Critical point.} Going down in $Y$, we locate the first step at
which the largest multiplier reaches $|\rho|=1$, the result is $Y_c(X)$.
\end{enumerate}

When the limit cycle is unstable, the Floquet multipliers also provide information about the number of memories switches before decaying to the silent fixed point. Assuming an initial condition with a non zero component $\epsilon$ only along the unstable eigenvector with eigenvalue $\bar{\rho}$, then after k full periods this components has grown to $\epsilon |\bar{\rho}|^k$. The alternation is destroyed once the unstable perturbation grows beyond the linear neighbourhood of the limit cycle and reaches a finite amplitude A, large enough for the trajectory to leave the vicinity of the cycle and enter the basin of attraction of a stable fixed point. This will happen after a number of periods given by the following equation:
\begin{eqnarray}
    k \simeq \frac{\ln (A/\epsilon)}{\ln(|\bar{\rho}|)}
\end{eqnarray}

\section{Memory-kernel timescale and equal time susceptibility}\label{app: mem_kernel_timescale}
\begin{figure*}[t]
\centering
\includegraphics[width=\linewidth]{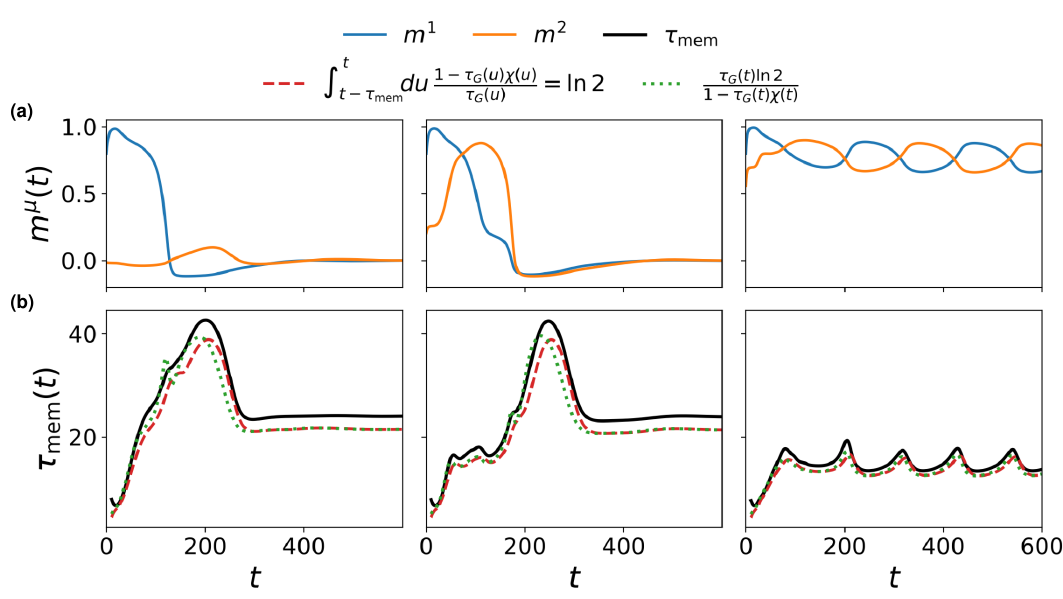}

\caption{Memory timescale $\tau_{\mathrm{mem}}(t)$ in the no-latching (left,
$X=0.22$, $Y=0.30$), high-latching (center, $X=0.80$, $Y=0.40$) and
pathological-latching (right, $X=0.86$, $Y=0.84$) regimes at $\alpha=0.1$.
Top: overlaps $m^1(t)$ and $m^2(t)$ with the two condensed patterns.
Bottom: $\tau_{\mathrm{mem}}(t)$ extracted directly as the half-width of the
memory kernel $\mathcal K(t,t')$ obtained from the self-consistent DMFT
solution, compared
with the two approximations of Appendix~\ref{app: mem_kernel_timescale}. Both
approximations use only two quantities extracted from the response function:
the equal-time response $\chi(t)=\tilde G(t,t)$ and the decay time of the
response, $\tau_G(t)$. Red dashed:
solution of
$\int_{t-\tau_{\mathrm{mem}}}^{t}du\,[1-\tau_G(u)\chi(u)]/\tau_G(u)=\ln 2$,
Eq.~\eqref{eq:taumem_exact}. Green dotted:
$\tau_{\mathrm{mem}}(t)=\tau_G(t)\ln2/[1-\tau_G(t)\chi(t)]$,
Eq.~\eqref{eq:app_taumem}, obtained by assuming that $\chi$ does not vary over
a time window of order $\tau_{\mathrm{mem}}$. Parameters: $S=3$, $a=0.25$,
$U=0.3$, $\beta=5$, $w=0.8$, $\tau_r=10$, $\tau_\theta=100$, $\Delta t=1$,
$m^1(0)=0.8$.}
\label{fig:taumem_tauG}
\end{figure*}

For $\alpha>0$, the single site dynamics has a non trivial memory kernel $K(t,t')$ that couples different colors of the same unit. For a unit of class $g$ and active color
$k=1,\dots,S$,
\begin{align}
\tau_r\,\dot r^k_g &= -r^k_g + H^k_g - \theta^k_g + \tilde h^k_g ,
\label{eq:r}\\
\tau_\theta\,\dot\theta^k_g &= \sigma^k_g - \theta^k_g ,
\label{eq:theta}\\
H^k_g &= \eta^{1}_{gk}m^1+\eta^{2}_{gk}m^2
 + w\Big(\sigma^k_g-\frac1S\sum_{l=1}^S\sigma^l_g\Big)+\frac{1}{\tilde a} \int_0^t dt'\sum_{l=1}^{S} C_{kl}
   \mathcal{K}(t,t')\sigma_g^l(t') +\phi^k(t),
\label{eq:H}\\
\sigma^k_g &= \frac{e^{\beta r^k_g}}{Z_g},\qquad
Z_g=e^{\beta U}+\sum_{l=1}^S e^{\beta r^l_g}.
\label{eq:softmax}
\end{align}
Here, $\tilde h^k_g(t)$ is an auxiliary external field, set to zero at the end.
The memory kernel is a functional of the response function of the system:
\begin{equation}\label{eq:app_mem_kernel}
        \mathcal{K}(t,t') = \alpha\sum_{n\geq 1}\tilde G^n(t,t'), \quad \tilde G(t,t')=\frac{1}{
\tilde{a}
}\sum_g c_g\sum_{kl}C_{kl}\,
\left.\frac{\delta\sigma^k_g(t)}{\delta h^l_g(t')}\right. .
\end{equation}
From simulations, we see that the response of the system is peaked at equal times, and it decays for $t'<t$. We define $\tau_G$ as the time scale over which it decays as follows
\begin{equation}
    \tilde G(t,t') =\chi(t)\,e^\frac{-(t-t')}{\tau_G}\,\Theta(t-t'), \quad \text{with} 
    \quad \chi(t) \equiv \tilde G(t,t).
\end{equation}
If we further assume that the system is in a stationary state and $\chi(t) = \chi$ is independent of time, then $\tilde G(t-t')$ depends only on the time-difference, and the product of two response functions is a convolution,
\begin{equation}
(\tilde G \tilde G)(t,t')=\int d u\;\tilde G(t-u)\,\tilde G(u-t')
=\int d v\;\tilde G\big((t-t')-v\big)\,\tilde G(v)=(\tilde G*\tilde G)(t-t').
\end{equation}
By induction,
$\tilde G^{\,n}(t,t')=\tilde G^{*n}(t-t')$, where $\tilde G^{*n}$ is the $n$-fold convolution.
Using the Laplace transform we can turn the convolutions in product and find the sum of the series that define the memory kernel.
The Laplace transform of $\tilde G(t-t')$ is:
\begin{eqnarray}
    \hat G(s) = \int_0^\infty du\; \tilde G(u)e^{-su}=\chi\int_0^\infty d u\,e^{-(s+1/\tau_G)u}
=\frac{\chi}{s+1/\tau_G}.
\end{eqnarray}
The convolution theorem gives $\widehat{\tilde G^{*n}}=\hat G^{\,n}$. Summing the
geometric series we can get the Laplace transform of the memory kernel $\hat K(s)$,
\begin{equation}
    \hat{\mathcal{K}}(s) = \alpha \sum_{n\geq1} \hat G^n(s) = \alpha \frac{\hat G(s)}{1-\hat G(s)} = \alpha \frac{\chi}{s+(1-\tau_G\chi)/\tau_G},
\end{equation}
and the inverse Laplace transform of $\hat{ \mathcal{K}}(s)$ is given by
\begin{equation}
    \mathcal{K}(t-t')=\alpha \chi\; \exp\Bigg[{-(t-t')\frac{1-\tau_G \chi}{\tau_G}}\Bigg].
\end{equation}
By defining $\tau_{mem}$ as the half-width time scale of the kernel, we obtain:
\begin{equation}\label{eq:app_taumem}
    \tau_{mem}(t)= \frac{\tau_G \ln2}{1-\tau_G \chi(t)}.
\end{equation}
Even though we derived this formula by assuming that $\chi$ does not depend on time, it is possible to show that we only need it to be a slowly varying function on time scales of order $\tau_{mem}$.
Indeed, if we do not assume $\chi(t) = \chi$, we can still find a relation between $\tau_{mem}$ and $\chi(t)$.
Define the integrated susceptibility
\begin{equation}
\Lambda(t,t')=\int_{t'}^{t}d u\;\chi(u),
\qquad\text{so that}\qquad
\partial_u\Lambda(u,t')=\chi(u),\quad \Lambda(t',t')=0 .
\label{eq:Lambda}
\end{equation}
If we assume that $\tilde G(t,t') =\chi(t)\,e^\frac{-(t-t')}{\tau_G}\,\Theta(t-t')$, we can then prove the following result 
\begin{equation}
\tilde G^{\,n}(t,t')=\chi(t)\,e^{-(t-t')/\tau_G}\,
\frac{\Lambda(t,t')^{n-1}}{(n-1)!}.
\label{eq:Gn}
\end{equation}
The proof goes as follows. For $n=1$ it is true because $\Lambda^0/0!=1$. If we assume it to be true for $n$, then:
\begin{align}
\tilde G^{\,n+1}(t,t')
&=\int_{t'}^{t}d u\;\tilde G(t,u)\,\tilde G^{\,n}(u,t')
\nonumber\\
&=\int_{t'}^{t}d u\;
\chi(t) e^{-(t-u)/\tau_G}\;
\chi(u) e^{-(u-t')/\tau_G}\frac{\Lambda(u,t')^{n-1}}{(n-1)!}
\nonumber\\
&=\chi(t) e^{-(t-t')/\tau_G}
\int_{t'}^{t}d u\;\chi(u) \frac{\Lambda(u,t')^{n-1}}{(n-1)!}
\nonumber\\
&=\chi(t) e^{-(t-t')/\tau_G}
\int_{t'}^{t}d u\;\partial_u\!\left[\frac{\Lambda(u,t')^{n}}{n!}\right]
\nonumber\\
&=\chi(t) e^{-(t-t')/\tau_G}\,\frac{\Lambda(t,t')^{n}}{n!}.
\end{align}
This is Eq.~\eqref{eq:Gn} for $n+1$.
Summing the series from Eq.~\eqref{eq:app_mem_kernel},
\begin{equation}
\mathcal{K}(t,t')=\alpha\sum_{n\ge1}\tilde G^{\,n}(t,t')
=\alpha\chi(t)e^{-(t-t')/\tau_G}
\sum_{m\ge0}\frac{\Lambda(t,t')^m}{m!}
=\alpha\chi(t)e^{-(t-t')/\tau_G}e^{\Lambda(t,t')} .
\end{equation}
and finally combining the
exponents,
\begin{equation}
\mathcal{K}(t,t')=\alpha\,\chi(t)\,
\exp\!\Big[-\frac1{\tau_G}\int_{t'}^{t}d u\,\big(1-\tau_G\chi(u)\big)\Big],
\qquad t>t'.
\label{eq:K_exact}
\end{equation}
$\tau_{mem}$ is then determined by the following equation:
\begin{equation}
\int_{t-\tau_{mem}(t)}^{t}d u\,\big(1-\tau_G\chi(u)\big)=\tau_G\ln2.
\label{eq:taumem_exact}
\end{equation}
It reduces to Eq. (\ref{eq:app_taumem}) if $\chi(t)$ varies slowly over a $\tau_{mem}$ window.
Fig. \ref{fig:taumem_tauG} shows $\tau_{mem}(t)$ extracted directly from the memory kernel, compared with the values obtained solving Eq. (\ref{eq:taumem_exact}), and using the approximate Eq. (\ref{eq:app_taumem}), using the value for $\tau_G(t)$ extracted directly from the $\tilde G(t,t')$:
\begin{equation}
    \tau_G(t) = \frac{\int_{-\infty}^t\tilde G(t,t')dt'}{\tilde G(t,t)}.
\end{equation}

\end{document}